\documentclass{aastex62}

\newcommand{\kms}{{km~s$^{-1}$}}
\newcommand{\masyr}{{mas~yr$^{-1}$}}

\newcommand{\moy}{$M_\odot$~yr$^{-1}$}

\newcommand{\nsample}{70}
\newcommand{\ngoodparallax}{231}

\shorttitle{Mass loss and bi-stability}
\shortauthors{Kobulnicky et al.}
\begin{document}

\title{MASS-LOSS RATES FOR O AND EARLY B STARS POWERING BOWSHOCK NEBULAE: EVIDENCE FOR BI-STABILITY BEHAVIOR }

\correspondingauthor{Henry A. Kobulnicky}
\email{chipk@uwyo.edu}
\author{Henry A. Kobulnicky}
\author{William T. Chick}
\affiliation{Department of Physics \& Astronomy, University of Wyoming, 
  Dept 3905, Laramie, WY 82070-1000, USA}
\author{Matthew S. Povich}
\affiliation{Department of Physics \& Astronomy, California State Polytechnic University, 3801 West Temple Avenue,  Pomona, CA 91768, USA}

\begin{abstract} 
Second only to initial mass, the rate of wind-driven mass loss determines the final mass of a massive star and the nature of its remnant.  Motivated by the need to reconcile observational values and theory, we use a recently vetted technique to analyze the mass-loss rates in a sample of OB stars that generate bowshock nebulae.  We measure peculiar velocities from new $Gaia$ parallax and proper motion data and their spectral types from new optical and infrared spectroscopy.  For our sample of \nsample\ central stars in morphologically selected bowshocks nebulae, 67 are OB stars.  The median peculiar velocity is 11 \kms, significantly smaller than classical ``runaway star'' velocities.  Mass-loss rates for these O and early B stars agree with recently lowered theoretical predictions, ranging from $\simeq$10$^{-7}$ \moy\ for mid-O dwarfs to 10$^{-9}$ \moy\  for late-O dwarfs---a factor of about 2.7 lower than the often-used Vink et al. (2001) formulation.  Our results provide the first observational mass-loss rates for B0--B3 dwarfs and giants---10$^{-9}$ to 10$^{-8}$ \moy.   We find evidence for an increase in the mass-loss rates below a critical effective temperature, consistent with predictions of the bi-stability phenomenon in the range $T_{\rm eff}$=19,000--27,000~K.  The sample exhibits a correlation between modified wind momentum and luminosity, consistent in slope but lower by 0.43 dex in magnitude compared to canonical wind-luminosity relations.  We identify a small subset of objects deviating most significantly from theoretical expectations as probable radiation-driven bow wave nebulae by virtue of their low stellar-to-nebular luminosity ratios.  For these, the inferred mass-loss rates must be regarded as upper limits.  
\end{abstract}


\section{Introduction} \label{sec:intro}

The rate at which massive stars expel material in radiation-driven winds is  a fundamental factor in their evolution.  Large mass-loss rates reduce final core masses  and, thereby, determine the type of supernova that ensues, as well as  the nature of the final compact object (e.g., white dwarf, neutron star, black hole, or none). For the $\simeq$50\% of massive stars that have close companions \citep{Kobulnicky2007,Sana2012,Kobulnicky2016}, mass exchange and common envelope evolution may become the overriding evolutionary influences.  But for single massive stars, stars with distant companions (effectively single), and  even in close binaries prior to interaction, wind mass loss removes copious amounts of envelope material on timescales relevant to the rapid evolution of such stars.  Measured mass-loss rates, $\dot M$, lie in the range  $<$10$^{-9}$ to few $\times$ 10$^{-6}$ \moy, depending on stellar mass and evolutionary stage \citep[e.g.,][]{Garmany1981,Howarth1989,Fullerton2006,Mokiem2007,Marcolino2009,Prinja2010}. \citet{Kudritzki2000}, \citet{Puls2008}, and \citet{Smith2014} provide comprehensive reviews of massive star winds and mass loss.

Measurements of mass-loss rates at any given spectral type and luminosity class span orders of magnitude. H$\alpha$, radio continuum, and infrared observations measure the excess above the stellar photosphere and constitute the class of ``$n^2$'' diagnostics,  since the extra-photospheric flux scales as the square of the density of material in the wind  for optically thin geometries. This class of techniques typically yields mass-loss rates at the upper end of the range and  larger than those predicted by theory unless corrected for the effects of ``clumping''---density inhomogeneities in the wind \citep[e.g., see ][for discussions of rates and clumping] {Owocki1988,Leitherer1988,Fullerton1996,Martins2005b,Puls1996,Puls2005,Massa2017}. The other canonical approach, ultraviolet absorption spectroscopy of high-ionization wind lines, typically yields mass-loss rates at the low end of the range and far below many theoretical models \citep{Garmany1981,Howarth1989,Fullerton2006,Marcolino2009}.  The former techniques become insensitive  below rates of about 10$^{-7}$ \moy \citep{Markova2004,Mokiem2007,Marcolino2009} while the latter becomes insensitive above  10$^{-7}$ \moy\ as many UV resonance lines become saturated.   In the limit of weak winds ($\dot M\lesssim10^{-8}$ \moy, $\log(L/L_\odot)\lesssim5.2$), UV-based mass-loss rates fall two orders of magnitude below theoretical  expectations.  This ``weak-wind problem'' \citep{Martins2005b,Mokiem2007,Marcolino2009,Muijres2012}, coupled with the limitations of the canonical mass-loss techniques, especially for a whole range of stars  later than about O7V, call for new approaches to diagnosing stellar mass loss.   Promising techniques include X-ray \citep{Cohen2014} or infrared \citep{Najarro2011} spectroscopy of stellar winds. 

 A related unsolved problem in stellar winds is the impact of the ``bi-stability'' jump, characterized by a sudden increase in Lyman continuum and metal line opacity over a narrow temperature range predicted variously to lie  somewhere between 19,000~K and 27,000~K\footnote{This is often termed the ``first bi-stability jump'', given the prediction of a ``second bistability jump'' near 9,000--12,000~K owing to a sudden recombination of \ion{Fe}{3} to \ion{Fe}{2} \citep{Vink1999,Petrov2016}.  In this work we consider only the former.} \citep{Pauldrach1990,Lamers1995,Vink2000,Petrov2016}, with modern estimates falling at the low end of this range.  This elevated opacity should  produce a dramatic increase in mass-loss rate by factors of three to as much as 20  and a corresponding decrease in the terminal wind velocity.   Observational verification of this putative bi-stable behavior is very limited.  \citet{Markova2008} observed a small sample of B supergiants on either side of the proposed bi-stability region and concluded that $\dot M$ increases only by factors of a few,  if at all.  Circumstantial evidence for larger enhancements in $\dot M$ comes from the slow rotation rates observed among B supergiants on the cool side of the bi-stability jump, interpreted as evidence for rotational braking through mass loss \cite[bi-stability braking,][]{Vink2010}. Our study provides new data on mass-loss rates that will have implications for the effects of bi-stable behavior in stellar winds.      

In \citet{Kobulnicky2018} we refined an underexploited mass-loss measurement strategy, building upon a principle outlined in \citet{Kobulnicky2010} and first proposed by \citet{Gull1979}.  The approach entails the physical principle of momentum flux balance between a highly supersonic stellar wind and the impinging interstellar medium around a high-velocity ``runaway'' \citep{Blaauw1961,Gies1986} star.   Along the surface where the momentum fluxes equate an arc-shaped shock (bowshock) forms \citep{Wilkin1996, Wilkin2000} that may be detectable in  the infrared continuum or emission lines (e.g., \ion{O}{3}, H$\alpha$). While only about 15\% of runaway stars have detectable infrared bowshock nebulae \citep{Peri2015}, over 700 candidate bowshocks around probable early type stars are known  \citep{Kobulnicky2016}. \citet{Kobulnicky2018} estimated interstellar densities preceding bowshocks and suggested that ambient number densities exceeding 5 cm$^{-3}$ may be required to produce a detectable infrared bowshock nebula.  This may explain why the Galactic latitude scale height of bowshock candidates \citep[0.6 degrees; ][Figure~3]{Kobulnicky2016} matches that of the molecular gas in the Milky Way.   Identified most commonly by their 24 $\mu$m morphologies, the majority of arcuate nebulae also have corresponding 70 $\mu$m nebulae \citep{Kobulnicky2017}.     

The mass-loss rates for such stars may be expressed in terms of quantities that are, in principle, observable:

\begin{equation}
 \dot M = { {4\pi R_{\rm 0}^2 V_{\rm a}^2 \rho_{\rm a}} \over {V_{\rm w}}} .
\end{equation}

\noindent  $R_{\rm 0}$ is the ``standoff distance'' between the star and the bowshock nebula, computed from an angular size on infrared images and a known distance. $V_{\rm w}$  is the terminal stellar wind speed, typically adopted from the literature according to the spectral type and luminosity class. $V_{\rm a}$ is the velocity of the star relative to the local interstellar medium (ISM), which may be computed from distance, proper motion, and radial velocity data in conjunction with a Galactic rotation curve model. Finally, $\rho_{\rm a}$ is the density of the ambient ISM, which \citet{Kobulnicky2018} estimated from the 70 $\mu$m infrared surface brightness using \citet[][DL07]{DL07} dust emission coefficients and assuming pre-shock/post-shock density ratio of 1:4 appropriate to strong shocks.  In convenient astrophysical units the mass-loss rate may be expressed as,  

\begin{equation}
 \dot M (M_\odot ~yr^{-1}) = 1.67\times10^{-28} { {   {[R_{\rm 0}{\rm (arcsec)}]^2} D{\rm (kpc)}~[V_{\rm a}{\rm (km~s^{-1})}]^2 ~I_\nu{\rm (Jy~sr^{-1})}} 
  \over {V_{\rm w}{\rm (km~s^{-1})} ~\ell{\rm (arcsec)} ~ j_\nu{\rm (Jy~cm^2~sr^{-1}~nucleon^{-1})}}} .
\end{equation}

\noindent Here, $j_\nu$ is the frequency-dependent dust emission coefficient per nucleon as a function of ambient radiant energy density parameter from the star's illuminating flux at the distance of the infrared nebula, $U$.\footnote{ $U$={$u_* \over u_{MMP83}$} = ${{R_*^2\sigma~T_{eff}^4/(R_0^2 c)}\over{u_{\rm MMP83}}} $ where $u_{\rm MMP83}$=0.0217 erg~s$^{-1}$ cm$^{-2}$/c,  given by \citet{Mathis1983}.}   $U$, in turn, may be obtained by knowing the standoff distance, the stellar luminosity from its effective temperature and radius, since the central star dominates the local radiation field by factors of 100 or more \citep{Kobulnicky2017}.  

Finally, $I_\nu$ is the observed infrared suface brightness at the corresponding frequency.  We use the surface brightnesses and scaled DL07 dust emission coefficients at 70 $\mu$m owing to the availability of {\it Herschel Space Observatory} ($HSO$) measurements in that band, but  other wavelengths may eventually be proven suitable.   We avoid using the 24 $\mu$m $Spitzer~Space~Telescope$ ($SST$) or 22 $\mu$m {\it Wide-Field~Infrared~Survey~Experiment} ($WISE$) bandpasses because of the likelihood that this wavelength regime may be dominated by a population of stochasitically heated very small grains \citep{DL07}. The path length through the nebulae is given by the observed angular chord diameter, $\ell$.  $D$ is the distance to the star and nebula.  

Our approach has the distinction of being rooted in a fundamentally different principle from other mass-loss measurement techniques. As such, we expect it to be insensitive to some of the uncertainties that limit other methods. Because the bowshocks typically lie several tenths of a parsec from the star,  small-scale effects of wind clumping or temporal fluctuations in wind speed or density should be minimized through spatial and temporal averaging.  At the same time, our technique is subject to its own set of assumptions and approximations.  
\begin{itemize} 
\item{We assume isotropic stellar winds.  But mass loss could be enhanced along the polar axis or reduced at the equatorial plane for rapidly rotating stars \citep{Owocki1996,Langer1998,Mueller2014}.}
\item{We assume a homogeneous interstellar medium. This is appropriate, on average, but certain to be incorrect for many real astrophysical examples. } 
\item{Our approach assumes that the velocity of the impinging ISM is directed radially in the frame of the star.  Bulk flows of interstellar material at oblique angles relative to a moving star have been considered theoretically \citep{Wilkin2000} and should produce asymmetric bowshock nebulae with morphologies similar to some of those cataloged in \citet{Kobulnicky2016}.  Such complications are beyond the scope of this effort. } 
\item{The 1:4 pre/post-shock density ratio we adopt to estimate the interstellar density preceding the shock is merely a physically motivated assumption appropriate to strong non-radiative shocks. Density ratios in regions that experience significant cooling in radiative shocks could be more extreme.  Hydrodynamical simulations of OB star bowshocks show that density ratios may reach factors of 10 over small scales that are well below the observational resolution limits. Factors of four are about right when averaged over $\simeq$0.1--0.3 pc scales typical of sample objects \citep[e.g., Figures 7 and 4, respectively, of][]{Comeron1998,Meyer2017}.  } 
\item{It should be remembered that the measured geometrical quantities such as $R_0$ are projected quantities only.  We apply a statistical correction factor of  1/sin(65\degr)$=$1.10 for inclination effects when computing $R_0$ in pc. This is a suitable correction because inclinations substantially smaller than about 50\degr\ would begin to mask the arcuate morphology and make the object unlikely to be included in the list of bowshock candidates (e.g., see \citet{Acreman2016} for numerical simulations of bowshocks at various inclination angles.)  Given the large variation in signal-to-noise ratios of infrared images, we have not attempted to derive inclination angles for each object using geometrical properties of the nebulae \citep[e.g., ][]{Tarango2018}.  }
\item{We assume that the DL07 dust emission coefficients are appropriate to bowshock nebulae where a combination of radiative and shock heating may exist.   In Section \ref{sec:coefficients} we describe a rescaling of these coefficients made necessary by the harder radiation field of early type stars comapred to the interstellar radiation field.  These dust models are constructed for dust that is radiatively heated.   \citet{Kobulnicky2017} found that the dust color temperatures of bowshock nebulae were systematically above those expected from steady-state radiative heating from the central stars, leading them to propose this difference as evidence for shock heating, although this signature could also result from stochastically heated grains (A. Li, private communication).   Additional heating by shocks could elevate dust emission coefficients beyond those adopted here.  If shock heating is present but not properly accounted for, the adopted dust emission coefficients could be too small and the resulting mass-loss rates too large (c.f., Equation 2).   To the contrary, \citet{Henney2019} argues that stellar wind particles do not propogate across the termination shock into the swept-up dusty region of the nebulae, and so shock heating is negligible.    }

\end{itemize}

\citet{Kobulnicky2018} employed the principle of momentum-balance for  20 bowshock-generating OB stars with known distances.  They found stellar mass-loss rates factors of 10 or more lower than canonical $n^2$ diagnostics for a homogeneous wind, factors of 10 greater than UV absorption-line determinations, and factors of $\approx$2 lower than recent theoretical mass loss predictions \citep{Vink2001,Lucy2010b}. They concluded that, once corrected for {geometrical clumping and porosity in velocity space}, theoretical models would produce $\dot M$ predictions consistent with the bowshock method measurements.  They further noted that the technique showed promise for measuring mass-loss rates for weak-winded late-O and early B stars, but there were not enough B stars in their sample to draw meaningful conclusions.  Furthermore, uncertainties on the \citet{Kobulnicky2018}  mass-loss rates were large and dominated by uncertainties on the stellar velocities (assumed there to be 30 \kms\ for lack of direct measurement)  and uncertainties on the distances to the stars.\footnote{The DL07 dust emission coefficients tabulated in \citet{Kobulnicky2018} are also affected by an interpolation error for objects with $U<10^4$.}  Because $\dot M$ in Equation~2 scales as the square of both $V_{\rm a}$ and $D$ in this technique ($D$, via the distance dependence implicit within the derived $U$ parameter on which dust emission coefficients are based), their results were highly sensitive to errors on these quantities.     

With the availability of the $Gaia$ mission Data Release 2 (GDR2) data products  \citep{Brown2018}, excellent parallaxes and proper motions---therefore, distances and peculiar velocities---are now available for an enlarged sample of bowshock-generating stars.  In \citet{Kobulnicky2016} we compiled an ``all-sky''\footnote{The visual search for arcuate 24 $\mu$m  nebulae was confined to regions near the Galactic Plane where both early type stars and mid-infrared surveys from the $SST$  are concentrated.}  catalog of 709 bowshock candidates consisting of arcuate infrared nebulae enclosing symmetrically placed stars.  In \citet{Kobulnicky2017} we presented mid-IR photometry for the catalog of 709 bowshock candidates using archival $SST$, $WISE$, and $HSO$ images.  In this contribution we extend the proof-of-concept sample of \citet{Kobulnicky2018} to measure mass-loss rates for stars having well-measured distances, velocities, and  infrared bowshock properties. This expanded sample includes a greater fraction of early B stars and represents the first attempt at mass-loss determinations for dwarfs in this stellar temperature and luminosity regime.  Section 2 describes the selection of sample targets and computation of requisite parameters.  Section 3 details the section of the sample of stars for analysis.  Section 4 presents the mass-loss rates with a comparison to previous literature results and theoretical model predictions.  Section 5 summarizes the implications for mass loss prescriptions and for the evolution of massive stars. 

\section{New Data}
\subsection{Infrared and Optical Spectra from the Apache Point Observatory}
We acquired new optical spectra of 15 candidate bowshock stars using the Double Imaging Spectrograph (DIS)\footnote{https://www.apo.nmsu.edu/arc35m/Instruments/DIS/} at the Apache Point Observatory (APO) 3.5-meter telescope on the nights of 2018 May 12, May 19, and June 13.   The 1200 line mm$^{-1}$ gratings in both the red and blue arms of the spectrograph yielded reciprocal dispersions of 0.58 and 0.62 \AA\ pix $^{-1}$, respectively, over wavelength ranges 5700--6900 \AA\ and 4200--5500 \AA, respectively.  Exposure times ranged from 2$\times$300 s to several$\times$600 s depending on source magnitude, yielding spectra with signal-to-noise ratios (SNR)  between 30:1 and 100:1 at 5900 \AA.   Seeing ranged between  1\farcs6 and 4\farcs0 in a 1\farcs5 $\times$110\arcsec\ slit aligned at the parallactic angle, owing to the large airmass for the southern targets.   The instrument HeNeAr lamps supplied periodic wavelength calibration to an RMS of 0.006 \AA\ in the red channel and 0.1 \AA\ in the blue channel.  Instrument rotation produces wavelength shifts of up to 0.3 pix during the night which were removed by periodic arc lamp exposures so that the wavelength solutions are estimated to be precise to about 5 \kms\ based on repeated observations of the same star.  On nights with good seeing where the FWHM of the point spread function was comparable to the slit width, placement of the star within the slit may contribute additional velocity uncertainties.  Observations of one or more radial velocity standard stars (HD~196850, HD~185270, and HD~182758) indicate that the velocity calibration is accurate to about 10 \kms\ in the worst cases.  We assign a minimum radial velocity uncertainty of 6 \kms\ to each star.   Data reductions employed internal quartz lamp flat fielding and local sky subtraction adjacent to the star.  Continuum normalization produced one-dimensional spectra suitable for spectral classification and radial velocity measurement after velocity correction to the Heliocentric reference frame using the IRAF\footnote{\citet{Tody1986}; IRAF is distributed by the National Optical Astronomy Observatories, which are operated by the Association of Universities for Research in Astronomy, Inc., under cooperative agreement with the National Science Foundation.} {\tt rvcor} task.    

We observed 25 stars (four of these were also observed with the DIS optical spectrograph) using the TripleSpec \citep{Wilson2004}  infrared cross-dispersed echelle spectrograsph at APO  on the nights of 2017 July 13, 2017 August 30, 2017 Sep 01, 2017 Sep 08, 2017 October 09, 2018 May 29, 2018 June 3, and 2018 June 24. The spectrograph yields continuous spectral coverage between 0.95 and  2.4 $\mu$m at a resolution of about R$\simeq$3500 (85 \kms) using a 1\farcs1 slit aligned at the parallactic angle.  Four or eight 60--120~s exposures were obtained over an airmass range of 1.0--4.7 using a standard ABBA nod pattern  on targets ranging from $H$=7 to $H$=12 mag, yielding spectra  with signal-to-noise ratios (SNR)  between 20:1 and 80:1 in the H band near 1.6 $\mu$m. Seeing averaged 1\farcs2--1\farcs8. Reductions involved flat fielding using internal quartz lamps and wavelength calibration using night sky emission lines adjacent to the target star yielding (vacuum) wavelengths to a precision of about 0.7 \AA\ ($\simeq$10 \kms). Observations of three radial velocity standard stars HD~182758 ($V_\odot$=+2 \kms), HD185270 ($V_\odot$=$-$23 \kms), and HD196850 ($V_\odot$=$-$21 \kms) indicate that the radial velocities are accurate to within about $\pm$12 \kms. This low instrumental precision is a consequence of the fact that the infrared point spread function was sometimes smaller than the 1\farcs1 slit width, causing the star to wander in the dispersion direction within the slit.    We assign a minimum radial velocity uncertainty of 6 \kms\ to each star. Several A0V stars were observed over the range of target airmasses to aid in removal of the telluric absorption features. Reductions were performed using the {\tt APOTripleSpectool}  IDL package, a modified version of the {\tt SpeXtool} package \citep{Cushing2004}.  Spectra were then transformed to the Heliocentric velocity frame using the $IRAF$ {\tt rvcor} task and continuum normalized, treating the $JHK$-band portions of the spectrum separately.    

We have supplemented these new spectroscopic data with 19 stars observed in the red portion of the optical spectrum with the optical longslit spectrograph at the Wyoming Infrared Observatory ($WIRO$) \citep[to be reported in ][]{Chick2019}.   Table~\ref{tab:basic} lists the source of spectrosocopy for each target using a single-character code for optical data from $APO$ (O---24 instances), infrared data from $APO$ (I---16 instances), optical data from \citet[][(C---19 instances)]{Chick2019}, and spectral types adopted from the literature (L---23 instances).  Twelve objects have observations from more than one observatory or wavelength regime.  In all, we have our own spectra for 47 of the 70 stars.   The Appendix provides a summary of spectral classifications for each object.   

The optical and/or infrared spectra were used to classify the stars and measure radial velocities.  Most stars were only observed on a single night.  Eight stars have both optical and infrared spectra, allowing for a comparison of radial velocities between the two regimes.  The overwhelming  majority of the stars show \ion{He}{1} features---notably 5875.65 \AA\ or  2.1126 $\mu$m---indicative of O and early B stars hotter than about 15,000~K.  A few show \ion{He}{2} at 5410 \AA\ or 2.1885 $\mu$m, a signature of stars earlier than about 09.  The equivalent widths of the He lines and  (if available) the ratios of \ion{He}{2}/\ion{He}{1} allows us to designate a  spectral type  for most stars to within about one subtype by reference to stellar atmospheric models \citep[Tlusty;][]{Lanz2003} or the infrared spectral atlas of hot stars  \citep{Hanson1996}.  Stellar atmospheric features in our observed wavelength range are relatively insensitive to temperature for stars in the B1--B4 range, and observed spectral lines (mostly \ion{He}{1} and \ion{H}{1}) are greatly degenerate between temperature and gravity. Hence, spectral classifications are more uncertain in this regime.  The available spectral diagnostics are not especially sensitive to gravity, so, in most cases the luminosity class is not well constrained from the spectra alone.  However, the addition of parallax distances allows us to distinguish between dwarf, giant, and supergiant luminosity classes: at a given distance only one luminosity class is consistent with the 2MASS JHK photometric data, once interstellar reddening is removed.  We measure and remove interstellar extinction using the Rayleigh-Jeans $H$$-$4.5 $\mu$m color excess method of  \citet{Majewski11}.  Appendix A contains a brief discussion of the spectral classification, distance, and radial velocity (with notes on possible binarity) for each star in the sample.  

\subsection{Rescaled Dust Emission Coefficients \label{sec:coefficients}}
\begin{figure}
\plotone{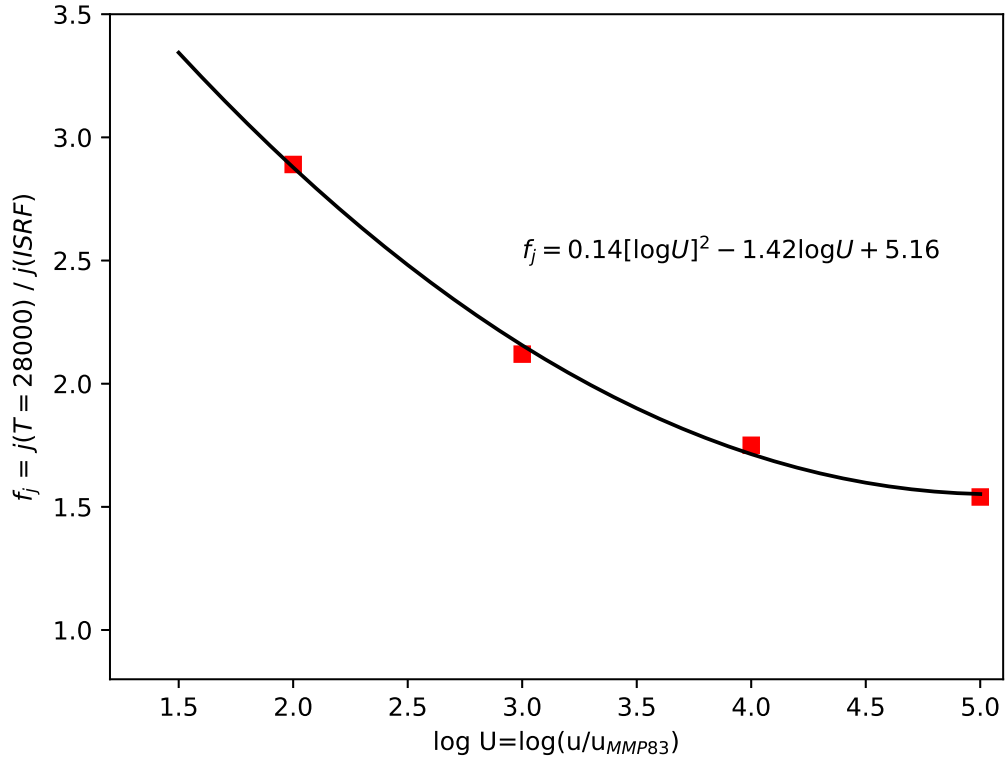}
\caption{Ratio of {\tt Cloudy}-derived dust emissivities, $f_j$, illuminated by a hot $T$=28,000~K source to illuminated by an interstellar radiation field having the same radiation density versus radiation density parameter, $U$.     
\label{fig:jratio}}
\end{figure}

\citet{Kobulnicky2018} used the dust emission coefficients of \citet[][DL07]{DL07} which are appropriate for  an incident radiation spectrum similar to the mean interstellar one \citep{Mathis1983}, peaking in the red or near-IR portion of the spectrum. However, the spectrum from OB stars illuminating bowshock nebulae is much harder, peaking in the ultraviolet.  \citet{Henney2019} concluded, on the basis of modeling an assortment of grain compositions using {\tt Cloudy} \citep{Ferland2017}, that it was necessary to scale the DL07 emission coefficients upward by about a factor of three across a range of $10^2<U<10^5$ in order to replicate the emission coefficients at the same value of $U$.  \citet[][Figure 4]{Henney2019} state that this is equivalent to using a DL07 model with a factor of about eight larger $U$.  In the absence of tabulated data, we performed {\tt Cloudy} modeling to compare the emissivity of standard ISM dust grains at 70 $\mu$m illuminated by a hot, $T$=28,000 K, $\log$ g=4.0 star \citep{Castelli2003} typical of our sample to emissivities illuminated by an interstellar radiation field ({\tt table ism} in {\tt Cloudy}) with the same radiation density.  We define this ratio of emissivities, $f_j\equiv j(T=28000)/j(ISRF)$.   Figure~\ref{fig:jratio} plots $f_j$ versus radiation density parameter over the range of our sample objects, $U$=10$^2$--10$^5$.  The black curve shows a polynomial fit to this relationship, $f_j(U)$=0.14[$\log$~U]$^2-$1.42[$\log$~U]$+$5.16.  At the low end of the range this ratio exceeds three, while at the high radiation densities it asymptotically approaches 1.5.  Varying the modeled stellar effective temperature from 18,000~K to 45,000~K, representing the full range in our sample, shows that this ratio varies by about 15\% across this wide range in temperature. Hence, a case-by-case, source-specific modeling effort would be appropriate as part of a future effort.  In this present analysis, we will adopt the $f_j(U)$ from Figure~\ref{fig:jratio} as a scale factor to increase the emission coefficients relative to nominal DL07 values to better approximate those expected in the UV-intense environments of bowshock nebulae . The net effect will be to decrease the derived mass-loss rates by factors of 1.5--3, with a median of 2.0.  The correction is more significant for objects with lower radiation densities.  These larger emission coefficients lead to reduced mass-loss rates compared to those in \citet{Kobulnicky2018}, and all values here supersede that work.  

We further note that our factor of 1.5--3 corrections to the DL07 dust emission coefficients, as indicated by Figure~\ref{fig:jratio}, are about 1.6$\times$ smaller than those adopted by \citet[][Figure~4]{Henney2019}, which rely solely on {\tt Cloudy} modeling.  In other words, a direct comparison of dust emissivities from DL07 and those derived from {\tt Cloudy} models for the same radiation density parameter and the same nominal interstellar radiation field and dust composition show that the former are a factor of about 1.6 smaller than the latter.  We are unable to identify a reason for this difference (A. Li, private communication). Hence, the absolute values of the dust emission coefficients should be regarded to entail a systematic uncertainty at the level of 60\% until systematic differences between DL07 and the dust implementation in {\tt Cloudy} can be resolved.  

%Volume emissivities in the 70 $\mu$m bandpass reported by {\tt Cloudy} in erg~s$^{-1}$~cm$^{-3}$ are converted to the DL07 emission coefficients in Jy~sr$^{-1}$~cm$^2$ nucleon$^{-1}$ by dividing the former by $4\pi n \delta\nu 10^{-23}$, where $n$ is the number density and $\delta\nu$ is the 70 $\mu$m bandpass width in Hz. 

\subsection{Distinguishing Bowshocks from Radiation Bow Waves}
\begin{figure}
\plotone{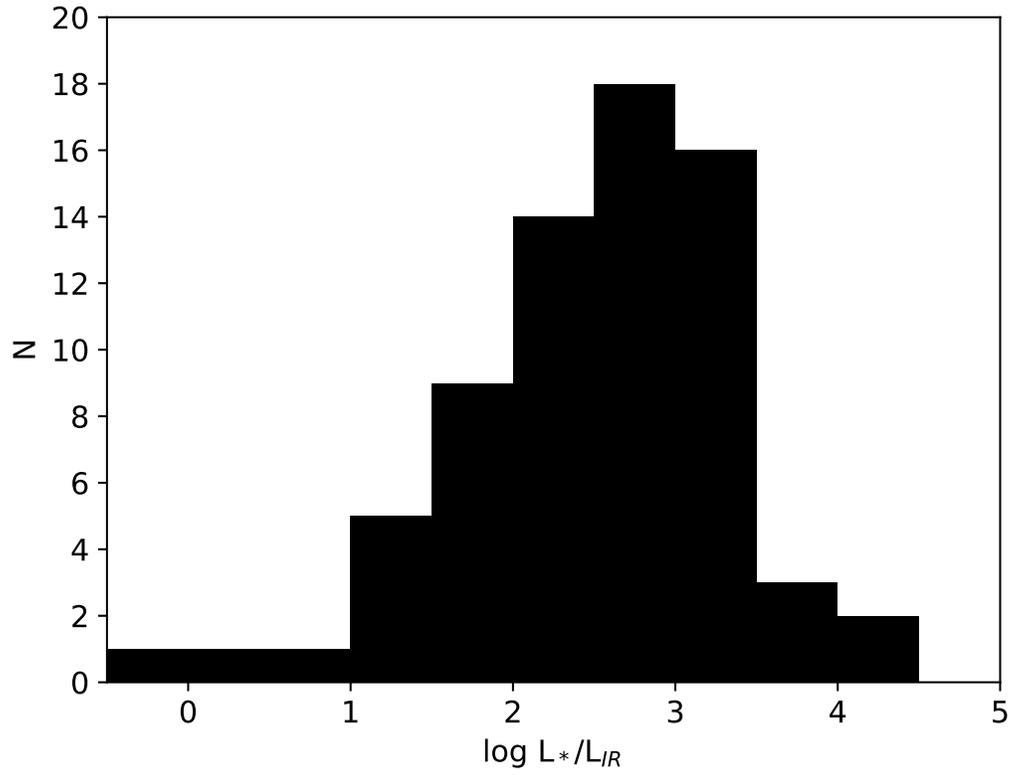}
\caption{Histogram of the logarithm of the stellar luminosity to infrared nebular luminosity for the 70 objects in this sample.  In objects with low values of $L_*/L_{\rm IR}$ the dust nebulae reprocess a significant fraction of the stellar luminosity, implying an appreciable optical depth for UV photons. The three lowest values are the three late-type stars in our sample.
\label{fig:LstarLIR} }
\end{figure}

\citet{Henney2019} articulated the physical distinction between stellar-wind-driven bowshocks and a class of ``radiation bow waves'' and   ``radiation bow shocks'' that may appear morphologically similar but originate from radiation rather than stellar wind pressure as the dust and gas in a nebula becomes optically thick.  Their Figure~2 distinguishes between these regimes as a function of three parameters: the stellar peculiar velocity, the standoff distance, and ambient interstellar density.  It appears that our hottest and highest-velocity stars typically fall into the regime of true wind bowshocks, but a few objects, particularly the cooler ones, fall into the ``radiation bow wave'' regime.  In the latter case, our derived mass-loss rates would become upper limits, as both stellar wind and radiation pressure play a role.    We make a crude assessment of the physical status of each bowshock using the ratio of stellar luminosity to infrared nebular luminosity, $L_*/L_{\rm IR}$.  This ratio ought to be large (e.g., $>$50) for true bowshocks where the dust optical depth is low and the nebula reprocesses a very small fraction of the stellar luminosity.  \citet{Henney2019} analyzed the 20 bowshock stars presented in \citet{Kobulnicky2018} and concluded that all but two were likely to be true bowshocks. In particular, object \#342 in the numeration of \citet{Kobulnicky2016} with $L_*/L_{\rm IR}$=22 is identified as an example of a trapped ionization front driving a radiation bow wave \citep{Henney2019}.   Figure~\ref{fig:LstarLIR} presents a histogram of $L_*/L_{\rm IR}$ for the sample.  The three lowest bins show the three late-type stars in our sample which have  $L_*/L_{\rm IR}<$10.  We will tentatively regard the objects with $L_*/L_{\rm IR}<$75 (15 stars) as candidate bow-wave nebulae, based on the analysis that will follow in Section~\ref{sec:massloss}.   The vast majority of the objects have $L_*/L_{\rm IR}>$75, making them likely to be genuine bowshock nebulae.

\section{Sample Selection for Measuring $\dot M$}   

Beginning with the \citet{Kobulnicky2016} table of 709 infrared-selected bowshock candidate stars, we searched the GDR2 for corresponding parallax and proper motion measurements. We found that 486 of the 709 stellar targets had a corresponding entry in the GDR2 within 1.5 arcsec (a few sources had two $Gaia$ entries within that error circle, and the brightest was taken as the most probable source).  However, we retained only the \ngoodparallax\ stars for which the parallax:uncertainty ratio exceeds 3:1, indicative of a reliable distance.  Of these \ngoodparallax, only 94 have bowshock nebulae with  secure $HSO$  70 $\mu$m detections required for estimating ambient densities \citep{Kobulnicky2017}.  Distances computed straightforwardly from inverse parallaxes are known to be systematically too large owing to the resulting asymmetry from parallax error uncertainties \citep{Bailer-Jones2015}. Although our 3:1 parallax:uncertainty selection criterion minimizes this bias, we adopt the distances from the Bayesian treatment by \citet{Bailer-Jones2018}, which results in 0\% -- 20\% (mean of 7\%) smaller distances compared to inverse parallax values. We also adopt as uncertainties their 68\% confidence limits.        

\begin{figure}
\plotone{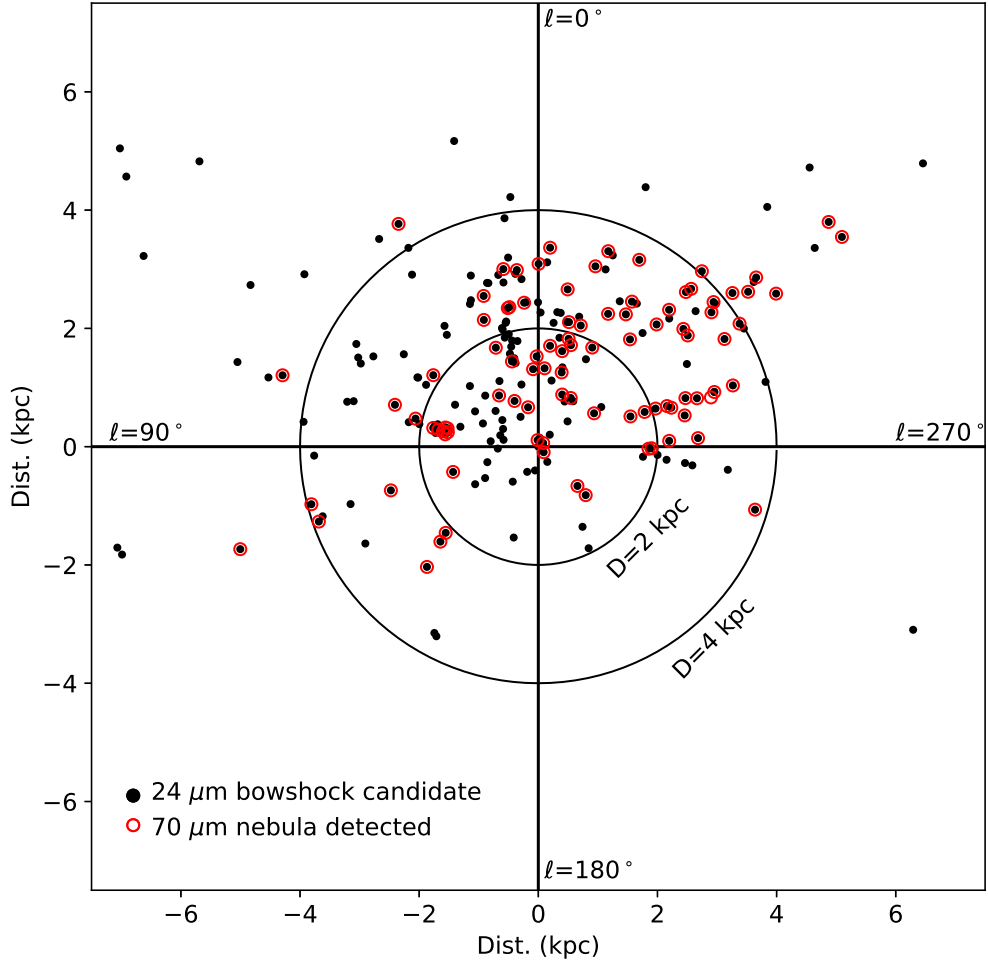}
\caption{Distribution of \ngoodparallax\ bowshock candidate stars having parallax:uncertainty ratios greater than 3:1 on the Galactic plane. The Sun's location is at the center of the Figure.  Red circles denote the subset of 94 that have 70 $\mu$m photometry from $HSO$. In this work we retain for analysis the 70 of these 94 that have measured spectral types.  
\label{fig:plane}}
\end{figure}

Figure~\ref{fig:plane} shows the locations on the Galactic plane of  the \ngoodparallax\ bowshock candidate stars with parallaxes known to 33\% or better ({\it filled circles}). The Sun is at the center. Concentric circles illustrate 2~kpc and 4~kpc distances. Larger unfilled red circles denote the subset of 94 stars having  detectable $HSO$ 70 $\mu$m nebulae.  Clusterings within Figure~\ref{fig:plane} reveal the concentrations of  bowshock OB stars in well-known Galactic associations such as the Cygnus-X region ($\ell\simeq79^\circ$; $D$=1.4--1.8~kpc) and the Carina complex ($\ell\simeq287^\circ$; $D$=2--3.5~kpc). The paucity of points in the second and third Galactic quadrants reflects the incomplete $SST$ 24 $\mu$m survey data in these regions, not necessarily a lower areal density of bowshock candidate objects.  The relative lack of sources beyond 4 kpc reflects the reduced sensitivity and angular resolution  of $SST$ surveys for distant targets in the bright, highly structured and confused mid-infrared images near Galactic mid-plane.  Many of the bowshock candidates tabulated in \citet{Kobulnicky2016} are small, near the angular resolution limit of the telescope; more distant sources would appear  point-like rather than arc-like and be excluded from the morphologically selected sample. Notably, nearly all of the objects in the fourth quadrant have 70 $\mu$m detections, while only about half in the first quadrant do. Finally, Figure~\ref{fig:plane} does not display obvious signs of spiral structure that might be traced by the OB stars as representative of young stellar populations.  However, the portion of the Galactic plane covered is small, and features more distanct than 4 kpc would not be seen using this tracer.  As noted by \citet{Xu2018}, distance uncertainties on OB stars in GDR2 make discernment of spiral structure problematic even for samples much larger than ours.  Furthermore, under the (yet-unvalidated and increasingly doubtful) hypothesis that bowshock stars are preferentially runaway stars, they may also fail to trace spiral structure if they have moved significantly from their birthplaces.   

Our intended $\dot M$ analysis also requires a measurement of the stellar effective temperature for the calculation of the dust emission coefficient, $j_\nu$.  This entails a secure spectral type for each target of interest.  Of the 94 star/nebula pairs having parallax \& proper motion data and 70 $\mu$m detections, we have collected spectral types for \nsample\ stars.  Our own optical and infrared spectra (as described above and in the Appendix)  provide the majority of these; a minority come from the literature.  A few of the tabulated targets are known or possible binary systems, noted as such in the Appendix.  Because multiplicity is high among massive stars \citep{Kobulnicky2007, Sana2012}, there are certain to be other undetected binaries among our sample.  Most of the new observations obtained for this work entail only a single epoch of spectroscopy, making binaries difficult to identify. Nevertheless, the near-absence of double-lined spectra in our sample suggests that, even in systems containing two or more stars, one star dominates the luminosity.  Notably, all but three of the \nsample\ sample stars show He features, confirming that they are, overwhelmingly, OB stars.  The remaining three non-OB stars turn out to be two (approximately) M giants  (\#129 and \#289 in the numeration of \citet{Kobulnicky2016}) and one K giant (\#653).  Whether these three late-type stars are responsible for the nearby arcuate nebulae or whether they are unrelated field stars mistakenly identified is unclear.  Ultimately, these are inconsequential, in a statistical sense, to the analysis that follows and are not discussed further.             

Table~\ref{tab:basic} lists basic data for the selected subsample of \nsample\ stars.  Column 1 contains the index number using the numeration of \citet{Kobulnicky2016}. Column 2 lists another common name of the star, followed by the generic name in Galactic coordinates in column 3.  Column 4 provides the adopted spectral type/luminosity class, primarily from this work and references cited herein (see Appendix).  Stars with an especially uncertain spectral type are designated by a colon. Columns 5 and 6 contain the adopted effective temperatures and radii, using the theoretical O-star temperature scale and radii from \citet{Martins2005a}.   For the few B stars we use the temperatures and radii of \citet{Pecaut2013}.   Column 7 gives the adopted stellar mass, again from  \citet{Martins2005a}. Column 8 provides the adopted terminal wind speed calculated by averaging Galactic O and B stars of the same spectral type from Table A.1 of \citet{Mokiem2007} and Table 3 of  \citet{Marcolino2009}.  An alternative approach to estimating terminal wind speeds based on the empirical calibration of \citep[][equation 9]{Kudritzki2000} yields velocities 1.5 times larger, on average.  The net effect of using these values would reduce derived mass-loss rates by the same factor, on average.  Given the different effective temperature scales that undergird that relation, we elect not to adopt these larger terminal wind speeds.   The adopted wind speeds are uncertain at the level of 30\%, based on the  dispersion at a given spectral type, and the fact that we do not have individual terminal wind speed measurements for these stars, which would obviously be desirable.  Early B giants and supergiants  are particularly uncertain with regard to both their wind speeds and stellar radii.  Column 9 lists the parallax-derived distances from \citet{Bailer-Jones2018}.  The distances used here are generally consistent with those pre-$Gaia$ distances used in \citet{Kobulnicky2018}, computed from other methods.  The largest differences concern the six objects assumed to be part of the Cygnus OB2 Association at the 1.32 kpc eclipsing binary distance \citep{Kiminki2015} but turn out to be much more distant, 1.7--1.9 kpc, according to $Gaia$ parallaxes. Column 10 lists the standoff distance, $R_0$, in arcsec, while column 11 lists the standoff distance in pc, calculated from  the distance in column 9 and the angular separation and the statistical projection factor of 1.1.  Column 12 lists the peak $HSO$ 70 $\mu$m surface brightness above adjacent background levels in 10$^7$ Jy sr$^{-1}$, occurring  at a location near the apex of the nebula.  Column 13 lists the angular diameter in arcsec of the nebulae along a chord ($\ell$) intersecting the peak surface brightness, as described in the text and Figure~1 of \citet{Kobulnicky2018}.   Column 14 lists the source of spectral classification: ``O'' for optical spectroscopy from $APO$, ``I'' for infrared spectroscopy from $APO$, ``C'' for optical spectra from \citet{Chick2019}, and ``L'' for literature spectral classifications.  Column 15 lists the ratio of stellar to infrared nebular luminosity, $L_*/L_{\rm IR}$, calculated from the stellar parameters, distance, and infrared measurements from \citet{Kobulnicky2017}. 

\begin{figure}
\plotone{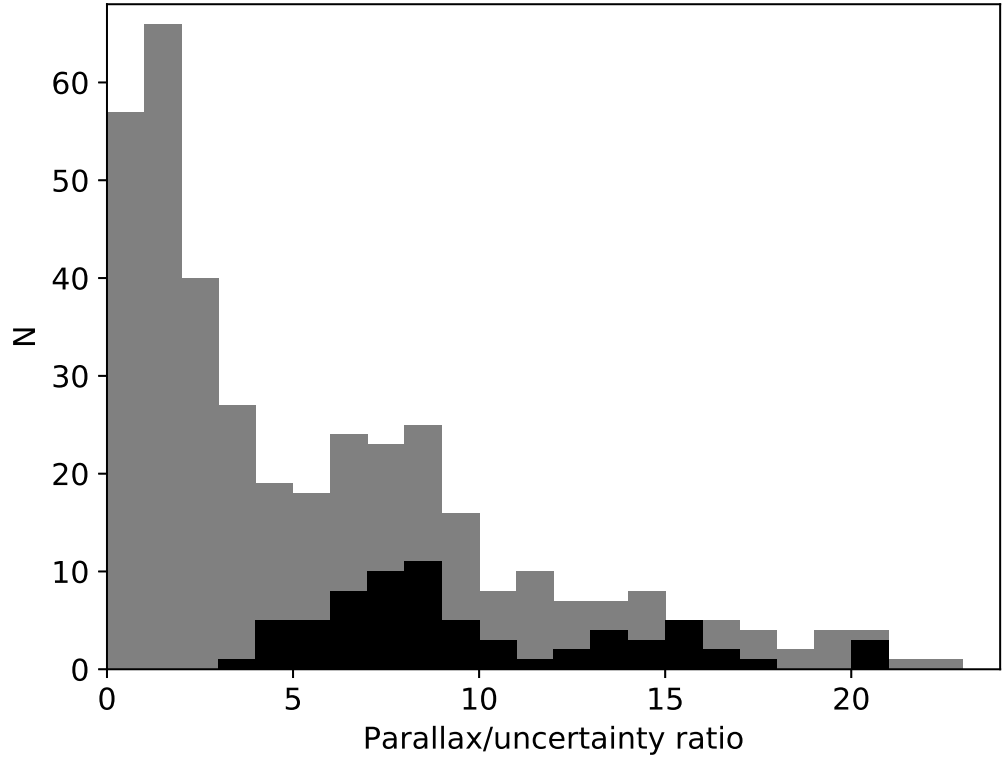}
\caption{Histogram (gray) of the ratio parallax/uncertainty for the 394 candidate bowshock stars with entries in the GDR2. The black
histogram shows the \nsample\ stars retained for further analysis.  These have known spectral types, 70 $\mu$m detections, and at least a 3:1 parallax:uncertainty ratio.  
\label{fig:parallaxes}}
\end{figure}

Figure~\ref{fig:parallaxes} is a histogram of the ratio parallax:uncertainty for the 394 candidate bowshock stars with entries in the GDR2 (gray shaded histogram).  Some GDR2 entries list negative parallaxes and are not shown.  Figure~\ref{fig:parallaxes} illustrates that the majority of bowshock candidate stars have parallax data that are (at present) insufficiently precise for reliable distance and proper motion calculation.  The black shaded  histogram shows the \nsample\ stars having known spectral types, 70 $\mu$m detections, and at least a 3:1 parallax:uncertainty ratio.  Thus, the typical star retained for analysis has a distance known to 15\% or better, which will reduce the uncertainties on $\dot M$ compared to those presented in \citet{Kobulnicky2018}.  

\begin{figure}
\plotone{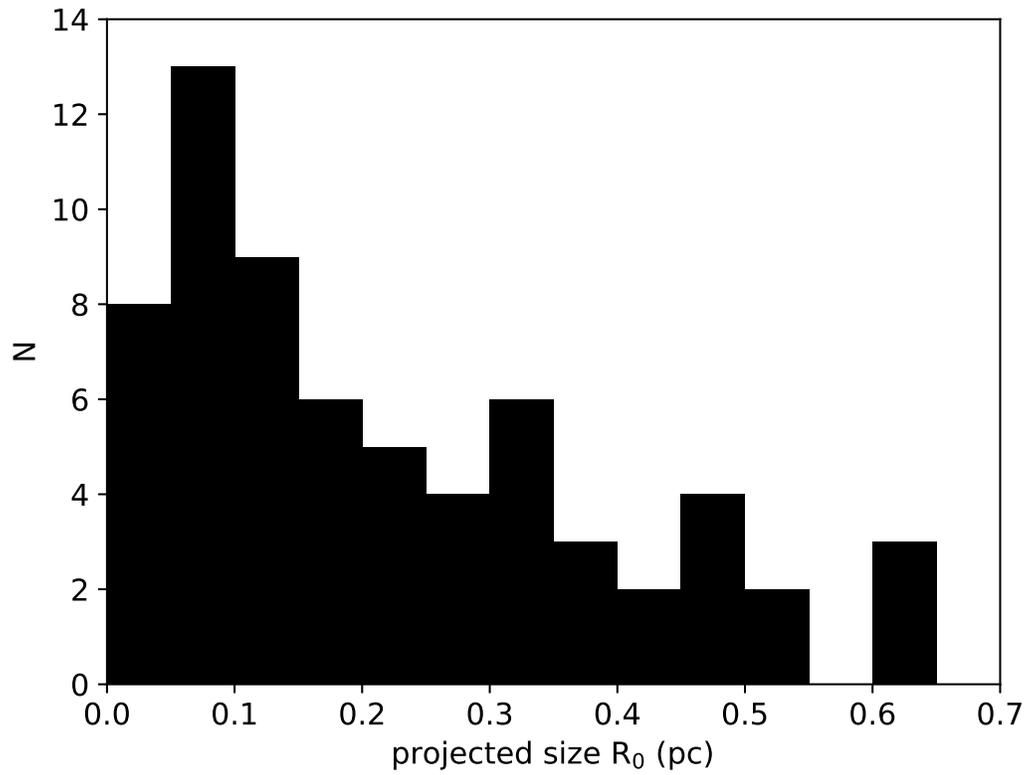}
\caption{Histogram of the {\it projected} standoff distances, $R_0$ in pc for the 70 objects with well-determined distances and secure 70 $\mu$m nebular detections.  The lowest bin is highly incomplete owing to the angular resolution limit of the $SST$ and $WISE$ observatories.  
\label{fig:standoff}}
\end{figure}

Figure~\ref{fig:standoff} presents a histogram of the {\it projected} standoff distances, $R_0$, in pc, for objects with well-determined distances and secure 70 $\mu$m detections.  The largest bowshock nebulae have characteristic sizes up to 0.6 pc, while 0.1--0.3 pc is typical.  The smallest bins are incomplete because of the angular resolution limit of the $SST$ images.  We expect that a 24 $\mu$m survey with arcsecond or better angular resolution would detect many additional arcuate nebulae, including objects at greater distances and objects with smaller standoff distances.    

After distances, peculiar velocity, defined as the deviation from the star's own local standard of rest, is the next most critical parameter.  Proper motion, radial velocity, and distance, used in conjunction with a model for Galactic rotation, are sufficient to calculate the velocity of the star relative to its own local standard of rest.  Ideally, precise radial velocity measurements would be available for each star.  Unfortunately, the radial velocity uncertainties on most of the objects are appreciable, a consequence of calibration and random errors totalling 6--12 \kms.  An even larger source of uncertainty is binarity.  At least 50\% of massive stars are found in binary systems where orbital velocities frequently exceed 100 \kms\ \citep{Kobulnicky2014}.   In order to prevent these sources of noise from dominating our estimate of peculiar velocities, we make the simplifying assumption that the stars' velocity vectors are primarily in the plane of the sky.  This is justified on the basis of the distinctive bowshock morphologies which would not be evident otherwise.  Accordingly, we set the radial velocity component of each star to zero in its local standard of rest and compute the peculiar velocities solely from the two orthogonal components inferred from GDR2 proper motions.\footnote{The one exception is $\zeta$ Oph, for which we use the superior data from $Hipparcos$ \citep{Perryman1997}.}  The reported peculiar velocities are, then, lower limits, but within 22\% (i.e., $\sqrt{3}/\sqrt{2}$) of those expected in the case of isotropic three-dimensional velocities. 

When computing peculiar velocities we apply the matrix transformation equations of \citet[][also see Appendix A of \citet{Randall2015} for implementation]{Johnson1987}. These transformations were applied assuming Galactic Center coordinates $\alpha_{GC}$~=~17h:45m:37s.224, $\delta_{GC}$~=$-$28d:56m:10s.23 and Galactic North Pole coordinates $\alpha_{GNP}$~=12h:51m:26s.282, $\delta_{GNP}$~=27d:07m:42s.01 \citep{Reid2004}. We adopt a solar galactocentric distance of 8.4~kpc and the Solar peculiar motion relative to the local standard of rest of (U,V,W)$_\odot$ = (11.1,12.2,7.2) \kms\ \citep{Schonrich2010}.  We adopt the Milky Way rotation curve of \citet[][Model I]{Irrgang2013}.  Our code for calculating peculiar velocities for bowshock stars uses the position, parallax, and proper motion for each star,  removes the peculiar solar motion (U,V,W)$_\odot$, computes the expected (U,V,W)$_*$ for the star's Galactic position given the adopted rotation curve, and computes the peculiar (U,V,W)$_{*pec}$ velocity of the star, i.e., the star's velocity relative to its local standard of rest.  Our code reproduces the velocities of over 1000 K--M dwarf stars \citep{Sperauskas2016} to within 1.1 \kms\ RMS. Uncertainties are propagated by Monte Carlo techniques.  Table~\ref{tab:kinematic} lists the identification number (column 1), generic name in Galactic coordinates (column 2), GDR2 identifier (column 3), observed Heliocentric radial velocity (column 4), parallax and uncertainty in microarcseconds (columns 5 and 6), observed proper motions and uncertainties in right ascension in $\mu$as yr$^{-1}$ (columns 7 and 8) and declination in $\mu$as yr$^{-1}$ (columns 9 and 10).  The radial velocities, which are certain to contain large contributions from unidentified binary orbital motion in some cases, are provided here for general interest only, and are not used further in this paper.  The calculated space velocities, assuming zero radial velocity components, are reported in Table~\ref{tab:derived}.   In eleven cases the calculated space velocity is very small---less than five \kms.  Given the documented bowshock morphology, the relative star-ISM velocity cannot be arbitrarily small.  Such sources may be examples of ``in-situ'' bowshocks \citep{Povich2008, Kobulnicky2016} caused by a bulk flow of ISM material.  We arbitrarily impose a minimum velocity of five \kms\ for these eleven objects.  

\begin{figure}
\plotone{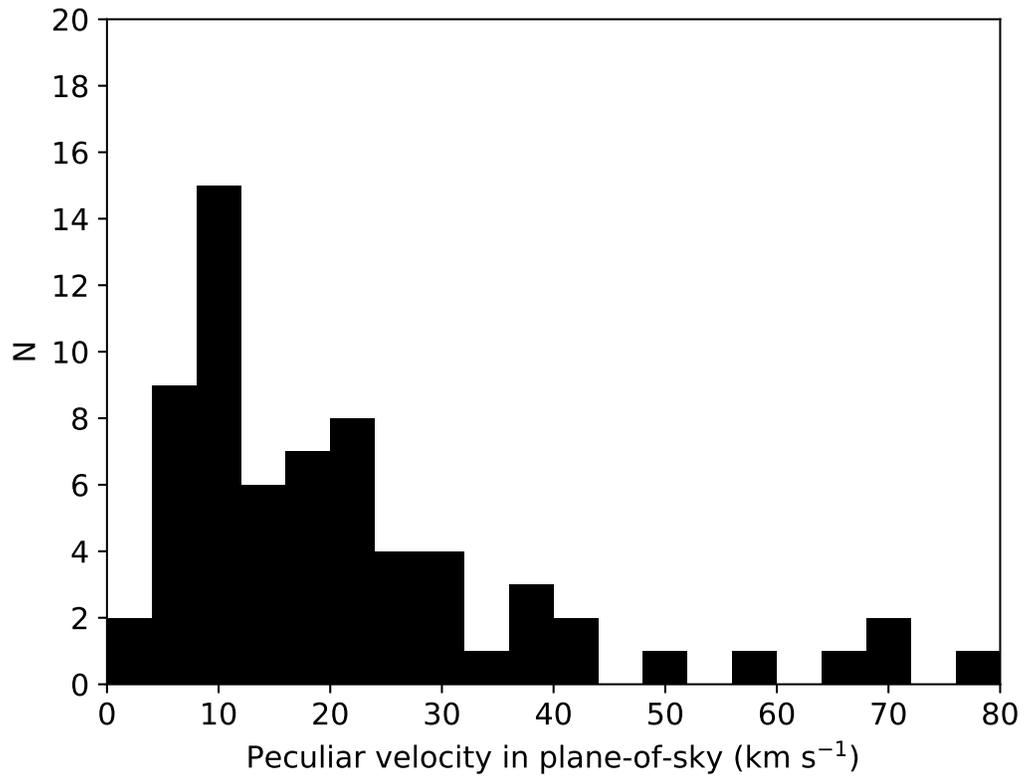}
\caption{Histogram of stellar peculiar velocities relative to their local rest frame.   The histogram depicts the \nsample\  stars selected for analysis.  
\label{fig:velocities}}
\end{figure}

Figure~\ref{fig:velocities} is a histogram of the stellar peculiar velocities, V$_{\rm tot}$.    Velocities range from near zero to 78 \kms.  The mean and median peculiar velocities for the \nsample\ sample objects is 15 and 11 \kms, respectively.  There are no extreme-velocity stars in the sample, which is not surprising since such star are not expected to produce visible bowshocks \citep{Meyer2016}  The mean peculiar velocity of 30 \kms\ assumed by \citet{Kobulnicky2018}, is, in retrospect, overly large.  However, there is a large dispersion of 16 \kms, so knowledge of each star's individual peculiar velocity is important in the computation of mass-loss rates, given the $V_{\rm a}^2$ dependence. Figure~\ref{fig:velocities} also reveals that the majority of bowshock-generating stars have a peculiar velocity less than 30 \kms\ and would not, on this criterion, qualify as ``runaway'' stars.  The possibility remains that some stars reside in regions of bulk ISM flows that introduce larger star-ISM relative velocities.  \citet{Kobulnicky2016} noted that an appreciable fraction of bowshocks appear oriented toward a neighboring \ion{H}{2} region where pressure gradients instigate ionized outflows and ISM velocities exceeding 30 \kms\ are plausible \citep{TenorioTagle1979,Bodenheimer1979}.  Figure~5 of \citet{Kobulnicky2016} shows four bowshocks oriented toward the ionizing sources in the M~16 nebula.  This is a good example of where bulk ISM flows may produce ``in-situ'' bowshocks \citep{Povich2008} that do not require high-velocity stars. 

\begin{figure}
\plotone{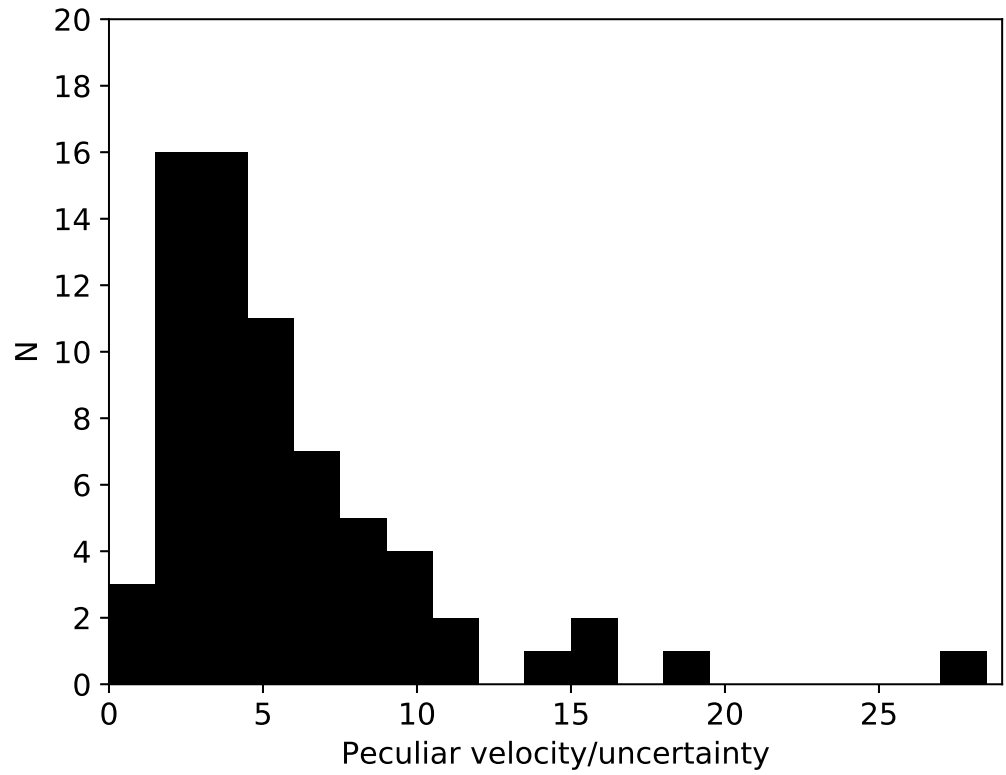}
\caption{Histogram of the ratio peculiar velocity:uncertainty.  The histogram depicts the \nsample\  stars selected for analysis.  
\label{fig:verr}}
\end{figure}

Figure~\ref{fig:verr} is a histogram of the ratio peculiar velocity:uncertainty, where the uncertainties are obtained by Monte Carlo error propagation.  The majority of stars have ratios less than 10:1. The median value for the \nsample-star subsample is 4.5, meaning that the typical velocity uncertainty is about 22\%.  Even with the new $Gaia$ data, velocity uncertainties remain significant when it comes to computing mass-loss rates, given the $V_{\rm a}^2$ dependence. 

\subsection{Computation of mass-loss rates} 

Table~\ref{tab:derived} contains quantities calculated from the basic data in Tables~\ref{tab:basic} and \ref{tab:kinematic}.   Columns 1--4 contain the identifying numeral, name, generic name, and spectral type, as in Table~\ref{tab:basic}.  Column~5 contains the stellar luminosity in units of 10$^4$ solar luminosities computed straightforwardly from the assigned effective temperature and radius.  Column 6 contains the radiation density parameter, $U$, calculated from the basic stellar effective temperature, radius, and standoff distance.   Column 7 lists the corresponding {\it scaled} 70 $\mu$m dust emission coefficient interpolated\footnote{We use a quadratic fit, $\log j_\nu =-0.066\log[U]^2+1.07\log[U]-14.19$, which is valid over the range $\log U$=1.5--5.0, and is superior to the linear interpolation used by \citet{Kobulnicky2018}, leading to slightly smaller emission coefficients in most cases. from the DL07 models in 10$^{-13}$~Jy~sr$^{-1}$~cm$^2$~nucleon$^{-1}$. We use the models for Milky Way dust in the $HSO$ 70 $\mu$m band with the minimum PAH contribution ($q_{\rm PAH}=0.47$\%) and single radiation density parameter ($U_{\rm min}$=$U_{\rm max}$) as suggested by the SED analysis in \citet{Kobulnicky2017}.  Column 8 is the factor by which the dust emission coefficients in Column 7 have been scaled from their original DL07 values, per Figure~\ref{fig:jratio}.} Column 9 is the  ambient interstellar number density, $n_{\rm a}$, derived from the 70 $\mu$m specific intensity, the bowshock cord length, $\ell$ and the adopted $j_{\nu}$, as described by \citet[][Equation 5]{Kobulnicky2018}.  Densities range between 4 cm$^{-3}$ and 2100 cm$^{-3}$, with a median value of 20 cm$^{-3}$. These are typical of densities within the cool neutral phase ($\approx$30 cm$^{-3}$) and the diffuse molecular phase ($\approx$100 cm$^{-3}$) of the interstellar medium and much larger than the warm neutral phase ($\approx$0.6 cm$^{-3}$) \citep[c.f.,][Table 1.3]{Draine2011}.   Column 10 lists the peculiar velocity, V$_{\rm tot}$, of the star derived from the proper motions, neglecting the radial velocity component. Column 11 gives its uncertainty.  We caution that this velocity is only the best attempt at measuring the velocity of the star relative to its local interstellar medium.  This velocity does not account for the possibility of local bulk flows of interstellar material, such as ``Champagne flows'' from a expanding \ion{H}{2} regions.  Columns 12 and 13 contain the mass-loss rates and corresponding uncertainties calculated from Equation 2.  Values range from 6$\times$10$^{-10}$ \moy\ to 5$\times$10$^{-5}$ \moy.   Column 14 lists the difference between the logarithm of our derived mass loss rate and the logarithm of the theoretical mass-loss rate, calculated using Equations 24 and 25 of \citet{Vink2001}.

In \citet{Kobulnicky2018} we assessed the uncertainties on $\dot M$ in terms of the uncertainties on each measurable parameter in Equation 2.   The angular quantities $R_{\rm 0}$ and $\ell$ are measured to about 10\% from infrared images, unchanged from our earlier work.  $I_\nu$ is measured to about 20\%, probably worse for some of the faintest nebulae. Mean stellar wind velocities are also unchanged, accurate to 30\% and showing real variation from star-to-star \citep[e.g.,][]{Mokiem2007}. Similarly, the uncertainty on $j_\nu$ is driven by the uncertainties on stellar temperature (taken to be 2000~K), radius (10\%), and standoff distance (we use the actual tabulated uncertainties), {however systematic uncertainties on the absolute values may exist at the 60\% level.}  Stellar distances and velocities---previously the dominant sources of uncertainty --- are now known much more precisely by virtue of the new $Gaia$ data, but still have significant uncertainties as shown in Figures~\ref{fig:parallaxes} and \ref{fig:velocities} and by the data in Table~\ref{tab:derived}.  We calculate uncertainties on the mass-loss rates using a 1000-iteration Monte Carlo error propogation procedure with the aforementioned uncertainties as inputs.  Our procedure would ideally start with the basic $Gaia$ parallax and proper motion measurements for a proper {\it ab inito} error analysis.  However, since we lack access to the \citet{Bailer-Jones2018} Bayesian code, we adopt the nominal distances and uncertainties from that work as well as the nominal stellar peculiar velocities and uncertainties they imply in order to further propagate the uncertainties from stellar temperature, radius, wind speed, and nebula angular quantities from Equation 2.  We find that a significant fraction of the iterations would result in unphysical negative peculiar velocities if the nominal velocity errors were used, e.g., 2$\pm$1 yields a non-negligible fraction of negative values.   This, combined with the likelihood that bulk ISM flows are an appreciable component of the star-ISM relative velocity, leads us to exclude velocity uncertainties from the Monte Carlo error analysis.  Therefore, the ensuing uncertainties on $\dot M$ appearing in the last column of Table~\ref{tab:derived}, should be regarded as indicative lower bounds.  The average uncertainty is 42\%  (excluding uncertainties on dust emission coefficients and stellar peculiar velocities). 

\section{Analysis } 
\subsection{Presentation of Mass-Loss Rates \label{sec:massloss}}
\begin{figure}
\plotone{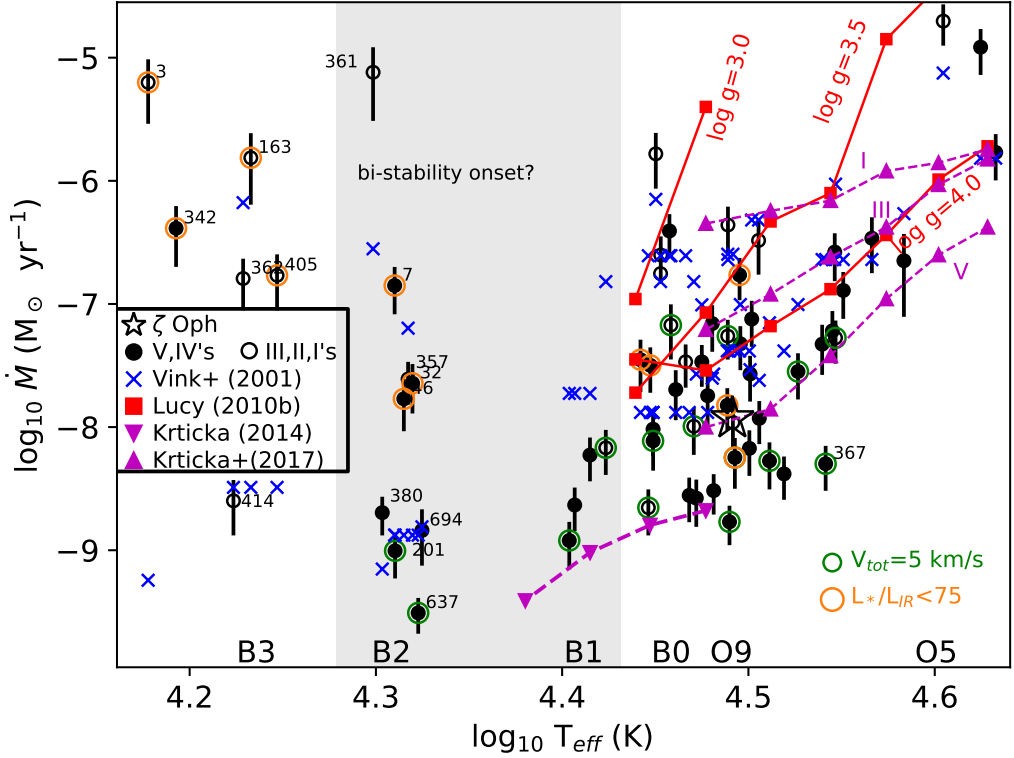}
\caption{Mass-loss rate versus stellar effective temperature.  Black filled points and black open circles depict measurements for main-sequence and evolved stars from this sample, respectively.  Blue crosses mark predictions for each object using the expressions of \citet[][Equations 24 \& 25]{Vink2001}.  Red squares and lines show the model predictions of \citet{Lucy2010b} for nominal main-sequence ($\log$ g=4.0), giant ($\log$ g=3.5), and supergiant ($\log$ g=3.0) stars, as labeled. The triangles and dotted lines show the theoretical predictions of \citet{Krticka2014} and \citet{Krticka2017} for B and O stars, respectively.  Green circles identify stars where the stellar peculiar velocity, $V_{\rm tot}$, has been arbitrarily fixed to the minimum values of 5 \kms.  Orange circles enclose objects that are candidate radiation bow wave nebulae. Two evolved B stars from Table~\ref{tab:derived} lie off the left side of this plot, as do the three cool late-type stars. A gray band marks the expected region of the bi-stability phenomenon.
\label{fig:mdot} }
\end{figure}

Figure~\ref{fig:mdot} plots the calculated mass-loss rates versus stellar effective temperature.\footnote{Although the mass-loss rate is expected to scale with {\it luminosity} rather than  {\it temperature}, (and to a lesser extent, metallicity; we only consider solar metallicities) we choose to plot the latter to facilitate direct comparison with the \citet{Lucy2010b} models.}    Black filled symbols denote the sample objects: a star for $\zeta$ Oph, filled circles for luminosity class V and IV stars,  and open circles for giants and supergiants.  A small dispersion has been added to the data points along the temperature axis to prevent marker pile-up.  Green circles enclose the objects where the peculiar velocities  were set to the minimum value of 5 \kms. Not surprisingly, these lie near the lower edge of the distribution.  Orange circles mark the objects having $L_*/L_{\rm IR}<$75 as potential radiation bow waves.  This criterion was selected based on the principles outlined in \citet{Henney2019} and after some experimentation that showed this cutoff selects {\it all} of the low-temperature stars with excessively large implied mass-loss rates but only a few of the hotter stars.   Two mid--late B stars from Table~\ref{tab:derived} lie off the left side of this plot, as do the three cool late-type (K--M) stars.  Blue crosses depict model predictions for each object using the  formulation of \citet[][Equations 24 and 25\footnote{We use the former for objects hotter than 27,000~K and the latter for cooler stars, given that we are unable to distinguish the correct branch  for objects in the 22,500~K--27,000~K transition regime.}]{Vink2001} computed using the stellar data from Table~1.\footnote{The downward revision of the solar metal abundance scale by \citet{Asplund2005} relative to the older \citet{Anders1989} scale used by prior works including \citet{Vink2001} would result in a predicted mass-loss rate lower by about 0.12 dex \citep{Krticka2007}. We have, accordingly, lowered all the pre-2005 theoretical predictions by this amount throughout this work. The correction is small compared to the dispersion in the data.}   We assume $v_{\infty}/v_{\rm esc}$=2 for stars on the hot side of the bi-stability jump and 1.3 for the cool side, per \citet{Vink2001}.  Hence, each filled data point is paired vertically with  a blue $\times$ at the same temperature, although the $\times$'s sometimes overlap.  Red squares connected by lines depict the model predictions from \citet{Lucy2010b} for (approximate) main-sequence ($\log$ $g$=4.0) and giant ($\log$ $g$=3.5) and supergiant ($\log$ $g$=3.0) stars, as labeled.  Magenta triangles and dashed lines show the predictions for B main-sequence stars from \citet[][solar abundance models]{Krticka2014} and O stars \citep{Krticka2017}, with separate tracks for stellar luminosity classes I, III, and V.     A gray band marks the predicted regime of the bi-stability jump.  

For the hot portion of the sample, the data show mass-loss rates rising from few$\times$10$^{-9}$ \moy\ near 27,000~K to over 10$^{-6}$ \moy\ for O5 stars near 42,000~K, but with a large dispersion of about half an order of magnitude. The data points for stars B1 and earlier fall systematically below the \citet{Lucy2010b} and \citet{Vink2001} predictions, but broadly consistent with---if slightly above---the \citet{Krticka2014} and \citet{Krticka2017} models.   The results are similar to those of \citet{Kobulnicky2018} but for a sample more than three times as large extending to cooler temperatures.  The dispersion of the data is larger than the typical uncertainties in the vertical dimension, consistent with additional uncertainty terms and/or real variations.  Considering the $\pm$1 subtype uncertainty on classifying any particular star, especially early B stars, any given object may fall into an adjacent temperature bin.   Evolved stars ({\it open circles})  generally lie above the luminosity class V and IV stars, also in accord with model predictions.  

The prototypical bowshock star $\zeta$ Oph, in particular, shows good agreement with the \citet{Krticka2017} models. With  $L_*/L_{\rm IR}=$1700, $\zeta$ Oph is safely out of the dust bow wave regime and should represent a good point of reference to studies employing other techniques.  Our nominal derived $\dot M$ of 1.2$\times$10$^{-8}$ \moy\ is a factor of 15 lower than the  estimate of 1.8$\times$10$^{-7}$ \moy\ from  \citet{Repolust2004}, which assumed no wind clumping correction, and hence, must be regarded as an upper limit.  Had we included the radial velocity component of $\zeta$ Oph's total space velocity (as one of the  stars where the assumption of zero radial velocity is known to be inconsistent with the data), the resulting space velocity of 26 \kms\ would imply a mass-loss rate of 5.5$\times$10$^{-8}$ \moy, still considerably lower than \citet{Repolust2004}, but only by a factor of three, which would be consistent with a standard correction for wind clumping.  Differences in adopted stellar parameters ($v_\infty$=1550, R$_*$=8.9 R$_\odot$, T$_{\rm eff}$=32,000~K versus 1300/7.2/31,000 adopted here) are relatively minor effects.  \citet{Repolust2004} notes that this star is a rapid rotator with $V_{\rm r}$$\sin$ $i$=400 \kms\ which could induce non-isotropic winds and modify the published values in either the observational or theoretical analyses, or both.  Our derived mass-loss rate for $\zeta$ Oph is still a factor of 10 or more larger than the  $\dot M$=1.5$\times10^{-9}$ \moy\ inferred from an analysis of X-ray line profiles \citep{Cohen2014} or the 1.6 $\times$10$^{-9}$ \moy\ from UV absorption lines \citep{Marcolino2009}.  Hence, there remains a considerable discrepancy between observational results on this prototypical ``weak wind'' late-O star.  

B1 and later spectral types in Figure~\ref{fig:mdot} fall considerably above the \citet{Krticka2014} prediction and above an extrapolation of the \citet{Lucy2010b} models.    These stars lie within the predicted regime of the bi-stability discontinuity.  They show a large dispersion in mass-loss rates from  $<$10$^{-9}$ \moy\ to almost 5$\times10^{-5}$ \moy, approaching --- but not exceeding --- the very large mass-loss rates predicted by \citet{Vink2018} for LBV stars, in a few instances.  Notably, a majority of the stars in this regime are candidates for being dust bow wave objects, having  $L_*/L_{\rm IR}<$75, making their derived mass-loss rates suspect.   However, the positions of some of these stars, particularly those that are not bow wave candidates, are broadly consistent with the \citet{Vink2001} prescriptions that take into account the effects of bi-stability below $\simeq$27,000~K. 

Figure~\ref{fig:vinkcompare} plots the difference of the logarithms between the mass-loss rates derived in this work and those predicted by the \citet{Vink2001} formulation versus effective temperature.  As before, filled markers denote (near-) main-sequence objects and open circles denote giants and supergiants.  A gray band highlights the range of the expected bi-stability jump.   Numerals adjacent to symbols identify each object, with a larger font for objects having our own optical spectroscopy and smaller italic font for other stars.  Black symbols denote comparisons performed using the \citet{Vink2001} upper/lower branch formulae for objects above/below 27,000~K, respectively.  Red symbols show stars having $T_{\rm eff}<$27,000~K {\it if the upper branch prescription  neglecting bi-stability effects were used instead.}   Green circles enclose objects where the lower limit peculiar velocity of 5 \kms\ was assigned.  Orange circles enclose candidate bow wave objects having $L_*/L_{\rm IR}<$75.  Blue squares enclose objects that directly face an 8 $\mu$m bright-rimed cloud where photoevaporative flows from the molecular cloud interface may influence the infrared nebulae.  Horizontal dashed lines mark the mean (bold dotted) and dispersion (non-bold dotted) for all points above 25,000~K.  The mean lies at $-$0.43 dex, indicating that, on average, the mass-loss rates are a factor of 2.7 below the \citet{Vink2001} prescription in in this regime, with a considerable dispersion of 0.64 dex (factor of 4).   A direct comparison with the predictions of \citet{Krticka2017} is not possible owing to the limited number of models tabulated in that work, but it may be concluded from Figure~\ref{fig:mdot} that our inferred mass-loss rates lie, on average, above those values by 0.1--0.3 dex.  The large dispersion in the data points in this figure results partly from the uncertainties on the derived mass-loss rate, typically 0.2--0.3 dex (see Table~\ref{tab:derived}, but recall that the tabulated uncertainties here do not include errors in $V_{\rm a}$),  systematic errors on dust emission coefficients, or additional uncertainties on the computed {\it theoretical} \citet{Vink2001} mass-loss rate (driven by uncertainties on the adopted parameters including the stellar luminosity, mass, effective temperature, terminal wind speed, and metallicity), amounting to another $\sim$0.2--0.3 dex.  

Figure~\ref{fig:vinkcompare} demonstrates that there is a systematic difference between our mass-loss estimates and theoretical models at temperatures $>$25,000~K and a much larger dispersion in mass-loss rates for stars $<$25,000~K.  The agreement between data and models is better when the prescription for bi-stability is applied---that is, the positions of black data points are consistent with the models minus a small offset versus the red symbols which depart dramatically from the predictions toward lower temperatures).  Unfortunately, our spectra do not contain the \ion{Si}{2} and \ion{Si}{3} lines in the blue portion of the optical spectrum required to assign temperatures to hot stars, so we are unable to ascertain the operative temperature where this deviation begins. Realistically, enhanced mass loss may become evident over a {\it range} of temperatures and luminosities depending on other factors traditionally linked to wind strength such as the stellar mass, metallicity, and rotation rate.  Indeed, \citet{Crowther2006} identified a systematic drop in terminal wind speeds across a 4000~K range in a sample of B supergiants but, these authors, as well as \citet{Markova2008} did not find a corresponding increase in mass-loss rates. Hence, our data in Figures~\ref{fig:mdot} and \ref{fig:vinkcompare} could represent the first evidence for enhanced mass-loss associated with the bi-stability phenomenon.  In particular, the most reliable data points (those {\it without} orange circles) show reasonable agreement with the \citet{Vink2001} predictions,  falling below the horizontal line by about 0.43 dex, similar to the hotter portion of the sample.  An extrapolation of the heavy dotted line marking the mean of the $T_{\rm eff}>25,000$ stars would intersect the most reliable B2--B3 objects in our sample, provided that the proper bi-stability formulation is used to predict the theoretical values.  Our estimate of the mass-loss rates for these objects, $\simeq10^{-9}$ \moy, for the four B2V stars represents the first results for dwarf stars in this temperature regime.  Their values are, like the hotter potion of the sample, about a factor of 2.5 below the \citet{Vink2001} predictions.      

If the especially high rates of mass loss for some cool objects were taken at face value, it ought to imprint an observable signature in the spectra in the form of extraphotospheric H$\alpha$ emission.  While we do not have high-resolution optical spectra of the sample stars that would be required to detect low levels of emission in the line cores, we do have low-resolution spectra for 37 stars.   Figure~\ref{fig:halpha}  plots the observed H$\alpha$ equivalent width versus stellar effective temperature for the sample stars with suitable data. Solid circles represent main-sequence stars and open circles represent evolved stars, as in previous figures.   Numerals designate each object by identification number. The red squares and lines mark theoretical values from model atmospheres \citep[Tlusty,][the O and B-star grids for solar metallicity]{Lanz2003} which are appropriate to plane-parallel static atmospheres without a wind.  Positive values represent absorption and negative values represent emission, per conventional usage.  While the equivalent width of H$\alpha$ is expected to increase from hot to cool (simply because Balmer lines becomes stronger toward cooler stars in this temperature range) across the bi-stability region, the data are mostly flat or decreasing, consistent with seeing elevated mass loss over the 27,000--19,000~K regime.   Most points in this range lie below theoretical expectations consistent with a  small contribution from  extraphotospheric emission. The lack of strong emission lines from the cool objects with very high implied mass-loss rates in Figure~\ref{fig:mdot} suggests that the infrared nebulae in such objects are driven by radiation pressure not winds, consistent with their small $L_*/L_{\rm IR}$ ratios. 

A comparison of outliers in Figures~\ref{fig:mdot}, \ref{fig:vinkcompare}, and \ref{fig:halpha} provides a means of assessing whether mass-loss rates are internally consistent.  Considering first the coolest objects at the left edge of both figures, the B4V \#342 and the B5III \#3 have derived mass-loss rates four orders of magnitude above the model predictions, even when bi-stability effects are considered.  Both objects have $H\alpha$ substantially in excess of the stellar photosphere, but not in emission, as might be expected for such large mass-loss rates.  While both objects have nebular surface brightnesses among the highest in the sample, they are not extraordinary in this regard.  But they are extraordinary in that the nebular surface brightesses are high for their late (cool!) spectral type.  Both nebulae are bright at 8 $\mu$m, indicating PAH emission consistent with a molecular cloud interface.  Furthermore, \#3 faces an 8 $\mu$m bright cloud rim \citep[][Figure 13.3]{Kobulnicky2016}, and is among the objects considered by \citet{Kobulnicky2016} to be affected by a photoevaporative flow from the inner edge of that rim.  The extent of the impacts from this phenomenon has not been theoretically developed.  Both \#3 and \#342  have  $L_*/L_{\rm IR}<75$, making them probable radiation bow waves.    Among the four evolved $\simeq$B3 stars, \#163 (B3II) and \#405 (B3III) lie 2--3 orders of magnitudes above \citet{Vink2001} predictions while \#362 (B2I) and \#414 (B3II) lie close to the model predictions.  H$\alpha$ for \#163 lies well in excess of the model photosphere (but not in emission).  This nebula lies along a 70 $\mu$m filament that does not show good morphological correspondence with the 24 $\mu$m nebulae.  It is likely that this object is confused by unrelated foreground or background emission.   In the case of \#405, we do not have H$\alpha$ data.  It faces the IC2599 \ion{H}{2} region associated with the energetic young cluster NGC~3324, which may drive an outflow and create a relative star-ISM velocity in excess of the tabulated 7.7 \kms. Both \#163 and \#405 are probable radiation bow wave nebulae.  Objects \#362 and \#414, on the other hand, agree well with model predictions once bi-stability behavior is considered. Neither object has evidence for being an instance of the radiation bow wave phenomenon. 

Stars in the B2 temperature range show mass-loss rate estimates in better agreement with model predictions, on average, but some still lie far above expectations in Figure~\ref{fig:vinkcompare}.  Object \#7 (B2V), for example, lies two orders of magnitude above \citet{Vink2001} predictions in Figure~\ref{fig:vinkcompare}.  Like, \#3, it is a probable radiation bow wave nebula.  It faces a bright-rimmed cloud prominent at 8 $\mu$m and may be affected by a photoevaporative flow \citep[][Figure 13.7]{Kobulnicky2016}. Examination of the mid-infrared image in \cite{Kobulnicky2016} shows another star 40\arcsec\ to the southwest of the nebulae that also lies along the nebula axis of symmetry.   Identified as an AGB star candidate \citep{Robitaille2008}, 2MASS J17582964-2610138 (R.A.(2000)=17h:58m:29s.64, Dec.(2000)=$-$26d:10m:13s.8) has H=10.4 mag and $A_{\rm K}$=3.08, an implied extinction that apparently precludes its inclusion in the GDR2, so its distance is unknown.  If this star is the source of luminosity powering the nebula it should be marked with an open circle in Figures~\ref{fig:mdot}  among other luminous evolved cooler stars.  This object should be regarded as uncertain in all respects.  Object \#361 (B1Ia; highly uncertain type and luminosity class) has a 70 $\mu$m nebula that is very near the detection limit, and the background is correspondingly very uncertain.  The derived mass-loss rate could be considered an upper limit.  Object \#32 (a probable B2V+B2V binary) lies near an extended 70 $\mu$m nebulosity that makes the background level somewhat uncertain. It is also a radiation bow wave candidate.  Object \#46 (B2V; also a radiation bow wave candidate) has an unusually large implied reddening for its parallax distance ($D_{\rm par}$=688$^{+163}_{-111}$ pc; $A_{\rm V}\simeq$11 mag; see Appendix), which may have implications for its nature and derived mass-loss rate.  Object \#380 (B2Ve) also lies faces a bright-rimmed cloud, a feature in common with other stars that show above-predicted mass-loss rates in Figure~\ref{fig:vinkcompare}.   The remaining four objects in this temperature regime (\#694--B2V, \#201--B2V, \#357--B2III, and \#637--B2V) all show mass-loss rates near the predicted values, and they all appear to be isolated objects with well-defined nebulae.  These objects comprise what we believe to be the first collection of B2 dwarfs with measured mass-loss rates, although two of these have the assigned minimum peculiar velocity of 5 \kms. 

The next group of three hotter stars near B1 shows improved agreement with the theoretical predictions.  All three lie slightly above the \citet{Krticka2014} prediction in Figure~\ref{fig:mdot} but substantially below the \citet{Vink2001} prediction, best visualized in Figure~\ref{fig:vinkcompare}.  Our data represent the first estimate of mass-loss rates for dwarfs of this temperature class.   

Other outliers in Figure~\ref{fig:vinkcompare} merit some discussion.  Object \#101 (B0V) faces an 8 $\mu$m bright-rimmed cloud and is a possible radiation bow wave with $L_*/L_{\rm IR}<75$ . Object \#626 (O9V) is also a radiation bow wave candidate, having the second highest nebular surface brightness in our sample.  Object \#339 (O5V) has one of the largest nebulae in our sample, and it has the largest peculiar velocity (78 \kms).  It also displays an H$\alpha$ EW considerably in excess of the photosphere in Figure~\ref{fig:halpha}. Object \#344 (the well-studied O4If star BD+43 3654) has the highest mass-loss rate in our sample (5$\times$10$^{-5}$ \moy), somewhat in excess of predictions.  There is the possibility that many of these objects are unrecognized binaries, and that their derived mass-loss rates should be reduced accordingly --- 0.3 dex for equal-luminosity components.   Objects that fall far below expectations are often those where the minimum peculiar velocity of 5 \kms\ has been assigned.  These include \#356 (B0III), \#463 (B0III), \#341 (O9V), and \#367 (O7V).  One other evolved object, \#407 (O9III+O), is a radiation bow wave candidate, but falls far below model expectations.   Object \#407 could also be an {\it in-situ} type bowshock as it does face directly toward the very energetic Carina star-forming complex at an angular separation of $\simeq$0\fdg2.  As such, its peculiar velocity may underestimate its true star-ISM velocity.  Conversely, object \#409 (O9.5IV) lies {\it above} model predictions and also faces the Carina nebula.      

\subsection{An Investigation of Biases by Subsample }

In an effort to investigate whether sample selection effects impact the essential conclusions from Figure~\ref{fig:mdot}, we considered only the half of the sample having nebular surface brightness above the median value.  We found that the essential conclusions from Figure~\ref{fig:mdot} are unchanged, with the O stars and early B stars showing good agreement with models and having a  similar dispersion as in the full sample.  However, fewer of the cooler stars are retained in this cut, indicating that the majority of the high-surface-brightness nebulae are, unsurprisingly, associated with hotter stars.  The lower surface brightness of the nebulae associated with the B stars in this sample means that they carry larger uncertainties, by virtue of the difficulty defining and subtracting local background emission around each nebula.  We also replicated Figure~\ref{fig:mdot} for the half of the sample having the highest velocity precision (indicated by peculiar velocity:uncertainty ratio $V_{\rm tot}$/$\sigma_V$)above the median value of 4.5.  The essential elements of  Figure~\ref{fig:mdot} remain.  O stars and early B stars show good agreement with the models and have a similar dispersion as before while the cooler stars exhibit the same dispersion observed in the full sample.  We conclude that including stars with larger velocity uncertainties does not bias the results in any particular direction.  Figure~\ref{fig:mdot1b} replicates the full set of objects from Figure~\ref{fig:mdot}, but magenta hexagons now enclose nebulae having 8 $\mu$m detections, suggestive of possible PAH contributions.  Sources with 8 $\mu$m detections lie scattered throughout the Figure, occurring both a low and high temperatures as well as at low and high mass-loss rates.  However, at the cooler end of the temperature range, most of the points that fall far above above the expected trend are those with PAH detections.  Many among this subset are also radiation bow wave candidates. Hence, at high temperatures the presence or absence of PAH features does not appear to bias the data in any particular direction, but at low temperatures it may be wise to treat such objects with caution.  

Table~\ref{tab:classifications} provides a list of object ID numbers that fall into each of the aforementioned special categories.  These are the 11 objects having $V_{\rm tot} <$ 5 km~s$^{-1}$, the seven objects that face 8 $\mu$m bright-rimmed clouds, and the 15 candidate bow-wave objects having $L_*/L_{\rm IR}<75$. 

\subsection{The Wind-Luminosity Relation \label{sec:WLR}}
 Figure~\ref{fig:MWM} plots the modified wind momentum \citep[][the product of mass-loss rate times terminal wind speed times square root of the stellar radius]{Kudritzki1995} versus stellar luminosity.  Filled circles denote luminosity class V and IV objects while open circles are luminosity classes I--III.  As in Figure~\ref{fig:mdot}, blue squares enclose objects that face an 8 $\mu$m bright-rimmed cloud, while green circles enclose objects where the minimum peculiar velocity was assigned, and orange circles enclose objects that are candidate radiation bow waves, $L_*/L_{\rm IR}<75$.  Because of the log scale, the uncertainties are similar for all data points. A single error bar denotes typical uncertainties which are dominated by the temperature uncertainty of 10\% (x-axis) and the uncertainties on the mass-loss rates (Table~3) and terminal wind velocities (y-axis).  The gray solid and dashed lines are theoretical relations for stars 27,500--50,000~K and 12,500--22,500~K, respectively \citep[][Equations 15 and 17]{Vink2000}, representing both sides of the nominal bi-stability jump.   

The dispersion among the data points is large, but a correlation is clear, consistent in slope with the theoretical wind-luminosity relation. The data fall systematically below this theoretical relation by about 0.4 dex.  Notably, the green circles fall exclusively on the low side of the trend and below model predictions.  Most of the objects that lie above model predictions are candidate radiation bow waves, and these often coincide with nebulae that face an 8 $\mu$m bright-rimmed cloud.   The most deviant objects in Figure~\ref{fig:mdot}, \#3, \#163, \#342, and \#361, also lie furthest from the theoretical relation in this figure.  Possible reasons for their departure from theoretical expectations have been discussed earlier.  The cool (B5--B8II, but highly uncertain) objects \#83 and \#391 appear on this plot but not on previous figures where they fall off the cool end of the axis range.  Both lie considerably in excess of the predicted relation, but neither is identified as a probable radiation bow wave nebula.  Both objects are cool enough that the second bi-stability jump near 10,000~K may be relevant.  Object \#361 (B2I) also lies well above the expected relation, but its temperature and luminosity is very poorly constrained by the available data.   Once the candidate radiation bow wave nebulae are removed, the remaining data in Figure~\ref{fig:MWM} show good agreement with the slope of the theoretical wind-luminosity relation, but with a zero point offset of about 0.4 dex toward lower values.   

\subsection{Comparison with Prior Data}

Figure~\ref{fig:mdot4} replicates Figure~\ref{fig:mdot} and includes mass-loss rates measured for the set of Galactic O3--O9.5 dwarf stars from \citet{Martins2005b} ({\it blue open triangles})  and \citet{Howarth1989} ({\it blue open squares}), respectively.\footnote{We have shifted the effective temperatures  assigned by \citet{Martins2005b} by $-$2000 K for consistency with the O9.5V objects in our sample.}   The \citet{Martins2005b} mass-loss rates derived from UV spectra fall somewhat below the bowshock sample, increasingly so in the regime of late O stars.   The mass-loss rates given by  \citet{Howarth1989} lie consistently above those in the present sample. The \citet{Howarth1989} values are broadly consistent with the \citet{Lucy2010b} theoretical expectations for dwarfs, but above the \citet{Krticka2017} relation.  Figure~\ref{fig:mdot4} also shows the late-O dwarf and giant stars  with mass-loss rates determined from ultraviolet P$^{4+}$ absorption lines \citep[][{\it green open and filled triangles, respectively}]{Fullerton2006}  and the same set of stars determined from the H$\alpha$ line \citep[][{\it cyan x's and +'s}]{Repolust2004,Markova2004}. The P$^{4+}$-based estimates show a large dispersion and lie consistently below the  mean of the bowshock sample, but there is some overlap.   The H$\alpha$ results  (for homogeneous winds) lie far above the mean of the bowshock stars by factors of ten or more. Some are upper limits, indicated by arrows, and  are consistent with the bowshock estimates.  This suggests that corrections for clumping in stellar winds are significant, factors of 3--10, consistent with inferences from other works \citep[e.g., review by][]{Puls2008}.  Unfortunately there are no objects in common (with the exception of $\zeta$ Oph) between the bowshock sample and the UV-based and H$\alpha$-based studies, so direct comparisons must await future works.  Overall, Figure~\ref{fig:mdot4} shows the bowshock sample agrees reasonably well with UV-based estimates, but lacks examples of extremely low mass-loss rates $\dot M < 10^{-9}$ \moy\ that typify the weak-wind phenomenon.  The bowshock sample also extends observational results into the regime of B0--B2 dwarfs that are not probed by other methods.  

\section{Discussion and Conclusions} 

\begin{figure}
\plotone{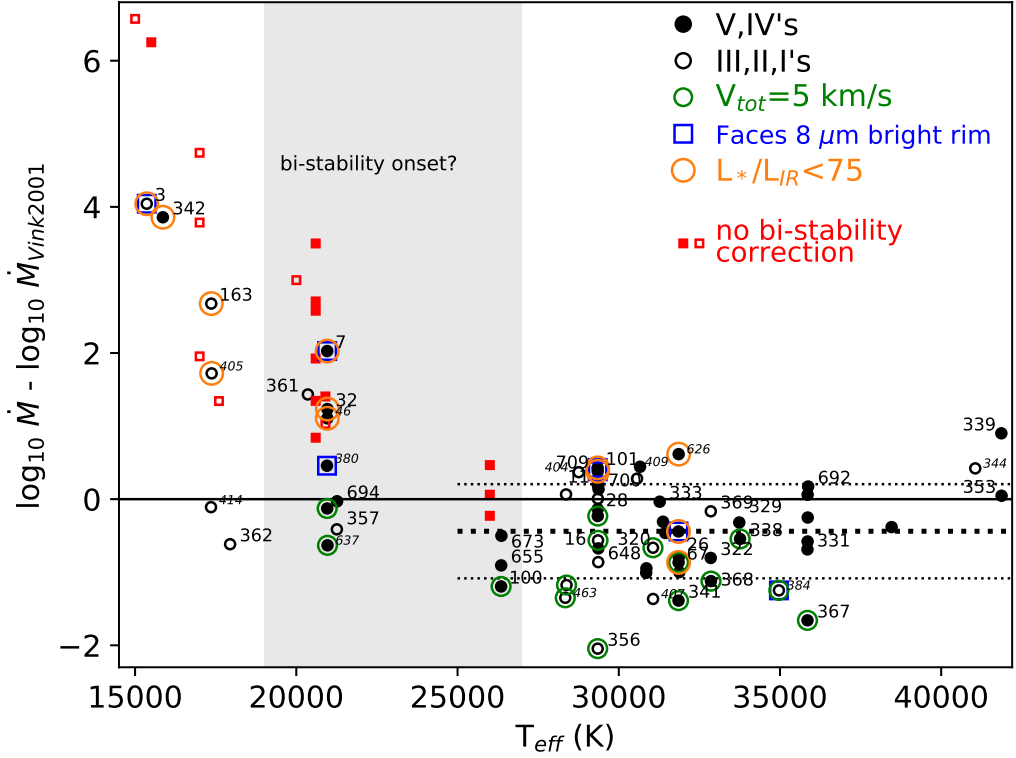}
\caption{The log of the mass-loss rate derived in this work minus the log of the \citet{Vink2001} prescription versus effective temperature.  Numerical labels identify stars having H$\alpha$ spectra presented in Figure~\ref{fig:halpha} (larger font) and other objects of interest (smaller italic font).   Black symbols denote comparisons performed using the \citet{Vink2001} upper/lower branch formulae for objects above/below 27,000~K, respectively.  Red symbols show stars having $T_{\rm eff}<$27,000~K if the upper branch prescription neglecting bi-stability effects were used instead.  Green circles mark objects where the lower limit peculiar velocity of 5 \kms\ was assigned. Orange circles enclose objects that are candidate radiation bow wave nebulae. Blue squares enclose objects that directly face an 8 $\mu$m bright-rimed cloud. 
\label{fig:vinkcompare} }
\end{figure}

\begin{figure}
\plotone{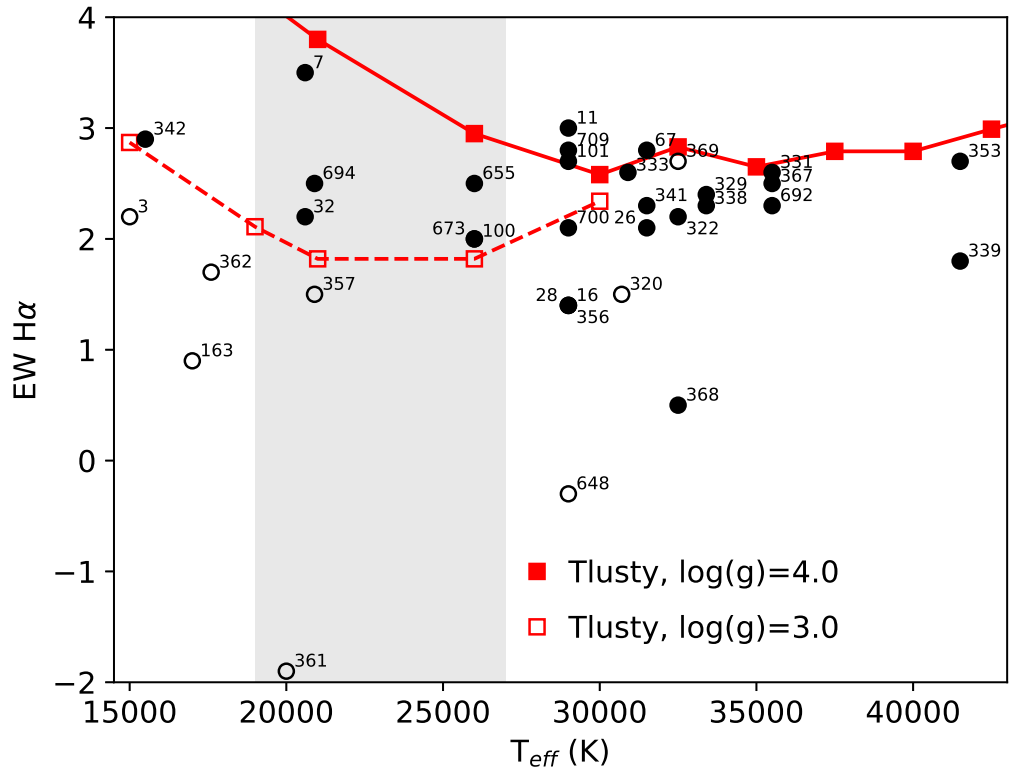}
\caption{H$\alpha$ equivalent width versus effective temperature for 37 stars with suitable data.  
The red points and lines are the expectations for dwarf and supergiant gravities from theoretical model spectra with  no mass loss \citep[][]{Lanz2003}.
\label{fig:halpha} }
\end{figure}

\begin{figure}
\plotone{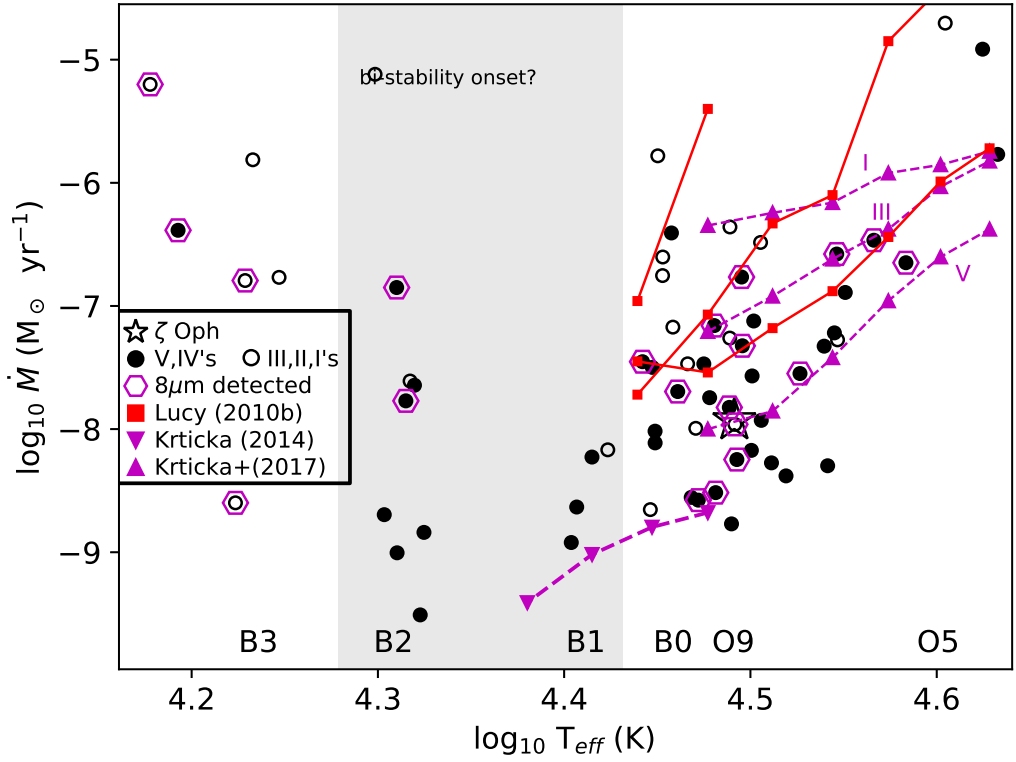}
\caption{As in Figure~\ref{fig:mdot}, but now objects with 8 $\mu$m detections, suggesting probable PAH contributions, are circled in magenta hexagons.  Objects with PAH emission lie scattered throughout the parameter space, indicating no significant bias on account of PAH-emitting nebulae.  
\label{fig:mdot1b} }
\end{figure}

Application of the physical requirement for momentum balance between a stellar wind and an impinging ISM yields a new method of mass-loss measurements for a sample of 67 early type stars.  The inferred relation between mass-loss rate and stellar temperature \& luminosity agrees well with several sets of theoretical models for main-sequence and evolved stars hotter than about 25,000~K, especially when models accounting for bi-stability behavior are employed, e.g., Figure~\ref{fig:vinkcompare}.  However, the mean mass-loss rates are lower by factors of about 2.7 compared to the canonical \citet{Vink2001} prescription, even allowing for a reduction of theoretical rates required by adoption of lower solar metallicity.  This factor of $\simeq$2.7 is consistent with the proposed reductions of 2--3 discussed in the literature and inferred from other measurement methods \citep[summarized by][]{Puls2015}. The derived mass-loss rates are in reasonable agreement with--but slightly larger than--the theoretical predictions of \citet{Krticka2017}.    Our data represent the first mass-loss estimates for dwarf stars in the very late-O and early B spectral ranges.  

\begin{figure}
\plotone{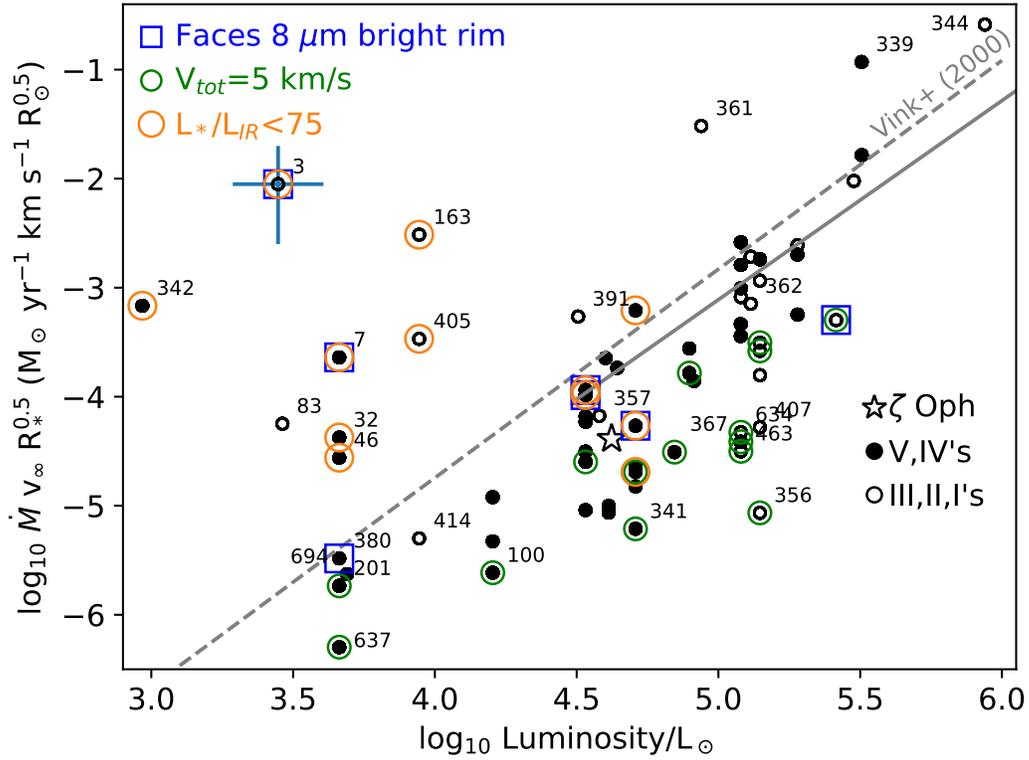}
\caption{Modified wind momentum versus stellar luminosity.  The gray solid and dashed lines denote the theoretical relations for stars 27,500--50,000~K and 12,500--22,500~K, respectively \citep{Vink2000}.  Filled circles denote luminosity class V and IV objects while open circles are luminosity classes I--III. Because of the log scale, the uncertainties are similar for all data points; a single error bar denotes typical uncertainties.        
\label{fig:MWM} }
\end{figure}

\begin{figure}
\plotone{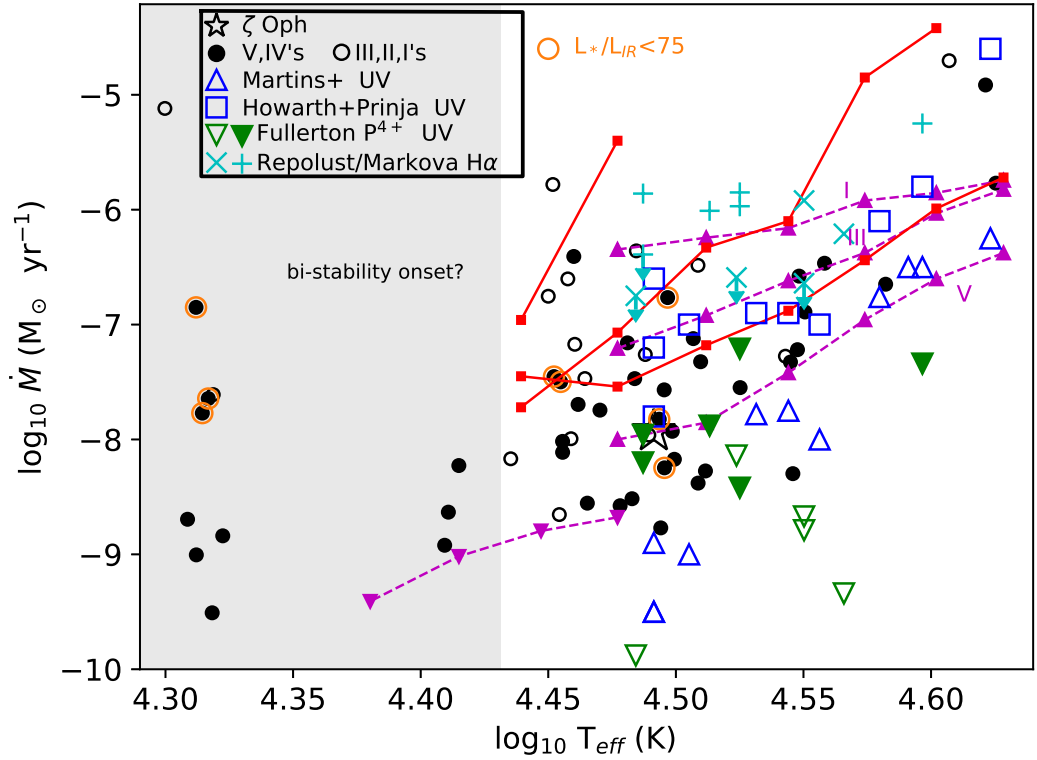}
\caption{Mass-loss rate versus effective temperature for dwarfs ({\it filled circles}) and evolved stars ({\it open circles}) from the bowshock sample.  Open blue triangles and squares depict observational results from the sample of Galactic O3--O9 main-sequence stars as measured by \citet{Martins2005b} and \citet{Howarth1989}, respectively. Green open and filled triangles depict the dwarfs and giants, respectively, measured using the ultraviolet P$^{4+}$ line  \citep{Fullerton2006}.  Cyan $\times$'s and $+$'s depict the same stars as determined from the H$\alpha$ line \citep{Repolust2004,Markova2004}.  Red and magenta lines mark theoretical predictions, as in previous figures.        
\label{fig:mdot4} }
\end{figure}

At temperatures cooler than 25,000~K the agreement is less good but is considerably better when bi-stability physics is applied rather than neglected, i.e., the black versus the red points in  Figure~\ref{fig:vinkcompare}.  We interpret this better agreement as a soft empirical confirmation that the prescriptions of \citet{Vink2001} get the temperature range and magnitude of bi-stability effects approximately correct.  Some of the cool objects in Figure~\ref{fig:vinkcompare} show excellent agreement with predictions while others show exceedingly high mass-loss rates compared to models.  Some of these large deviations may be explained by additional effects, such as when stars facing an 8 $\mu$m bright-rimmed cloud  experience an impinging photoevaporative flow which amplify the nebular surface brightness.  Five of the seven such objects in our sample (blue squares in Figures~\ref{fig:vinkcompare}, and \ref{fig:MWM}) show elevated mass-loss rates compared to predictions.   The majority of the most discrepant data are also probable radiation-driven bow waves where the nebula is becoming optically thick to UV photons.  In these cases, the inferred mass-loss rates can be regarded as upper limits.  Such objects should be discarded from efforts to measure mass-loss rates unless corrections can be developed.   In the few cases where the derived mass-loss rates lie below predictions, some of these objects face giant \ion{H}{2} regions where bulk outflows may produce a greater star-ISM relative velocity than we derive from $Gaia$ proper motions, leading to an underestimate of $\dot M$.  For some objects  $Gaia$ proper motions imply an unrealistically low peculiar stellar velocity, given the distinct arcuate nebulae observed, so we have assigned a minimum $V_{\rm tot}$ of five \kms. Ten of the eleven such stars show mass-loss rates  below the mean of the sample, leading us to consider these estimates of the mass-loss rates to be lower limits.  This sample, as a whole, exhibits the power-law relation  between modified wind momentum and luminosity very similar in slope---but not in zero point---to that predicted from classical wind momentum arguments (Figure~\ref{fig:MWM}).  Outliers on this Figure closely track the deviations observed in  Figure~\ref{fig:vinkcompare}.  The coolest objects show the largest discrepancies, and most of these outliers appear to be examples of radiation dust waves than true wind bowshocks \citep{Henney2019}.  Such objects will warrant careful investigation on a case-by-case basis. 

While the mean mass-loss rates agree well with recent theoretical expectations of \citet{Krticka2017} across much of the temperature range in the current sample, the dispersion is large, an indication of substantial sources of random error.  The tabulated uncertainties on $\dot M$ average 42\%. This does not include systematic uncertainties on dust emission coefficients or contributions from errors on the peculiar velocity---a very significant source of error owing to the velocity-squared dependence.  We expect this term to dominate the uncertainties and contribute greatly to the observed dispersion in the mass-loss rates, especially in light of the (unquantifiable) contributions from bulk flows in the ISM, with magnitudes 30 \kms\ or larger \citep{TenorioTagle1979,Bodenheimer1979}.  All of the other terms in Equation~2 which contribute to the sources of {\it random} error are well-constrained or contribute only linearly to $\dot M$.  We may also be seeing true variation in $\dot M$ at a given spectral type that results from second-order effects such as rotation velocity, metallicity, binarity, or stellar age.        

These new results are noteworthy for being the first mass-loss measurements in a sizable sample of early B dwarf and giant stars.  As a new physically independent technique for measuring mass-loss rates, our approach is thought to be unaffected by the same systematic effects (e.g., clumping, assumptions about dominant state of ionization) that have led to highly discrepant results from canonical techniques.  However, there are new sources of systematic error to consider.   Adoption of a slightly different calibration of stellar terminal wind speeds would lead to a systematic increase in wind speeds by a factor of 1.5 and a decrease in the derived mass-loss rates by the same factor.  Working in the opposite direction is a systematic increase of the stars' peculiar velocities; if we were to consider the full three-dimensional motions, the $V_{\rm tot}$ values could increase by as much as 22\%, on average, which would raise the derived mass-loss rates by a factor of 1.49.  There remains the possibility that bulk ISM flows generate larger star-ISM relative velocities than those measured from proper motions, potentially raising the derived mass-loss rates by significant factors for what we expect to be only a small subset of objects.   Additional systematic uncertainties on the dust emission coefficients at the level of 60\% need to be considered as well.  Adoption of updated physically specified dust emission coefficients surrounding hot stars, when available, could further revise the derived mass-loss rates.  On one hand, additional heating from wind shocks could elevate the dust emission coefficients beyond those used here, thereby reducing derived mass-loss rates.  Indeed, \citet{Kobulnicky2017} noted that the infrared dust color temperatures of bowshock nebulae exceeded those expected from reprocessing-of-stellar-photons arguments.  On the other hand, shocks could also destroy grains, thereby reducing dust emission coefficients.   We are unable to estimate the magnitude (or even the sign!) of a revision to the \citet{DL07} emission coefficients in bowshock nebula environments on the basis of existing work.  This issue remains, in our opinion, the largest source of systematic uncertainty on our approach to deriving mass-loss rates.  

Ultimately, a larger sample of bowshock-producing B1--B3 stars of all evolutionary phases need to be studied to understand the characteristics of stars that are typical of this population.  We expect an expanded sample will preferentially have small standoff distances and/or faint bowshock nebulae, and may not be plentiful in the present angular-resolution-limited samples.  Knowledge of the rotation rates (at least for B dwarfs) and evolutionary status of bowshock-producing B stars will also be important for a more complete understanding of these initial results.  One intriguing idea raised by the apparent success of the present results is to use the principles of bowshock physics outlined here to {\it measure} the speed of bulk ISM flows.  This would require that mass-loss rates are securely established through other observational or theoretical means.  Then, Equation 2 could be solved for $V_{\rm a}$ and, once peculiar stellar motions are removed, the bulk ISM speeds could be measured.  Stellar bowshocks could then truly become the ``interstellar weather vanes'' presaged by \citet{Povich2008}.  Future releases of the $Gaia$ data will provide improved distances and proper motions, reducing uncertainties on this sample of stars while opening up larger sample of bowshock stars to analysis.  Additional spectroscopy covering the blue/ultraviolet portion of the optical spectrum could provide much-needed stellar temperatures and wind speeds for each object in the sample.  While obtaining these data this remains an objective of our group's efforts, the faint magnitudes and high interstellar extinctions will necessitate substantial observational resources.   Theoretical work on dust emission coefficients that includes the role of stellar wind shocks as well as radiant heating would help place the present results on more secure foundation.

\appendix

This appendix includes discussion of individual objects based on optical or infrared spectra obtained at the Apache Point Observatory in 2017 and 2018 and optical spectra reported in \citet{Chick2019}. For each object we have matched the apparent central star to objects in the  GDR2 catalog.  For each star we provide the $Gaia$ identifier and the $Gaia$ (G) magnitude. We have also estimated spectrophotometric distances using the 2MASS $K$ magnitude, and we have adopted absolute magnitudes and colors from \citet{Pecaut2013}.  We correct for interstellar extinction using the Rayleigh-Jeans color excess prescription of \citet{Majewski11}.   Discrepancies between spectrophotometric distances and parallax distances are used to refine the luminosity class of stars which typically cannot be determined from the spectra alone.   Radial velocities measured from optical spectra are given in the Heliocentric frame and are based on the \ion{He}{1} 5875.75 \AA\ centroid (in-air wavelength).  Velocities from infrared spectra are based on the Br 12 line at  16411.55 \AA\  and Br 13 line at 16113.60 \AA\ (vacuum wavelengths), determined from PHOENIX model atmospheres (also given in the vacuum wavelengths) for hot stars \citet{Husser2013}.  Velocity uncertainties are limited by the SNR of the spectra and the broad nature of spectral features in massive stars, but are estimated at 6--11 \kms\, including systematic calibration uncertainties. Velocities, in most cases, result from a {\it single} optical or infrared spectral measurement; given the prevalence of binary systems with semi-amplitudes of hundreds of \kms\ among massive populations, these velocities should be taken as indicative but not a definitive measure of the systemic radial velocity.

G000.1169$-$00.5703 (\#1) --- Our infrared spectrum of this star (GID=4057291747277127040, $G$=10.7 mag) shows strong Br $\gamma$ emission and \ion{He}{1} 2.1126 $\mu$m in emission and \ion{He}{2} 2.1885 in absorption with a EW of about 1 \AA. The spectral features and photometric properties are consistent with an O8Ve star at the parallax distance of  $D_{\rm par}$=2441$^{+450}_{-332}$ pc. $V_{\odot}$=$-$15 \kms. 

G001.0563$-$00.1499 (\#3) ---  Our optical spectrum of this $G$=16.3 mag, 2MASS H=11.1 mag star (GID=4057634103373474560) observed at 2.1 airmasses is faint and shows H$\alpha$ but no He lines in our low-SNR optical spectrum.  Our infrared spectrum shows weak \ion{He}{1}.  Its  parallax distance ($D_{\rm par}$)of 1527$^{+242}_{-184}$ pc would be consistent with its spectrophotometric distance ($D_{\rm spec}$) if it were a B5III, so we have adopted this classification.  There are no other candidate stars at the geometric center of the arcuate nebula.  The H$\alpha$ velocity is $V_\odot$=$-$39 \kms\ which agrees roughly with the $-$15 \kms\ measured from the H lines in our infrared spectrum, so we adopt $-$27 \kms. The large proper motion in declination of $-$12 mas yr$^{-1}$ is much larger than any other star within 30 arcsec and is consistent with the bowshock orientation, making this a strong candidate for the source of the bowshock nebula.  The nebulae appears projected inside a larger broken ellipse of bright 8 $\mu$m PAH emission and has PAH emission at the location of the 24 $\mu$m arcuate nebula.   

G003.8417$-$01.0440 (\#7) ---  Strong \ion{He}{1} lines are present in the optical and infrared spectra of this $G$=12.7 mag object (GID=4064254301560852352), but no \ion{He}{2} are evident. B2V fits the spectrum well, but leads to a $D_{\rm spec}$=2000 pc --- slightly larger than $D_{\rm par}$=1313$^{+89}_{-78}$  pc.  The radial velocity computed from optical He lines is -0.5 \kms, in exact agreement with the radial velocity derived from the infrared H lines.  The $SST$ images shows additional nebulosities preceding the  24 $\mu$m  bowshock, suggesting an interaction with a more extended interstellar structure. No other $Gaia$ sources are located within 10\arcsec, but there are a collection of other stellar sources 20\arcsec\ to the southwest, so there may be other contributions to the ultraviolet flux that illuminates this region containing several arcuate objects.  

G005.5941+00.7335 (\#11) ---  B0V (GID=4067804899497964416; $G$=13.9 mag)  fits the He and H lines in the optical spectrum well, predicting $D_{\rm spec}$=2400 pc versus $D_{\rm par}$=2446$^{+437}_{-325}$ pc.    $V_\odot$=+18.8 \kms. Lines are especially narrow, suggesting the possibility of an evolved luminosity class, but our optical spectrum is not of sufficient quality to discern. The positive declination proper motion of 1.4 \masyr\ contrasts with other nearby objects, which are negative, and is consistent with the bowshock orientation.    
 
G006.2812+23.5877 (\#13) --- $\zeta$ Oph (GID=4337352305315545088, $G$=2.4 mag)  is the prototypical runaway bowshock star, an O9.2IV.  We adopt the radial velocity of $V_\odot$=$-$15 \kms\ and calculate is peculiar velocity from its HIPPARCOS parallax (distance of 112 pc), proper motion, and radial velocity rather than the $Gaia$ data which carry larger uncertainties.  

G006.8933+00.0743 (\#16) --- B0 provides a good match to the optical spectrum this star (GID=4069437463777653632, $G$=11.7 mag). $D_{\rm par}$=3008$^{+801}_{-534}$ is consistent with $D_{\rm spec}$=2760 pc only if the luminosity class is near III.   $V_{\odot}$= $-$24 \kms.  

G011.0709$-$00.5437 (\#26) --- A probable O9V (GID=4094704481502396288, $G$=11.7 mag) this star shows weak \ion{He}{2} 5410 \AA\ in our optical spectrum.  It has  $D_{\rm par}$=3060$^{+471}_{-364}$ pc, somewhat larger than  $D_{\rm spec}$=2280 pc, but this discrepancy could be reconciled if it were an equal-luminosity binary.   $V_{\odot}$= $-$1 \kms. 

G011.6548+00.4943 (\#28) --- This star (GID=4095751942430811776, $G$=13.2 mag) is approximately a B0V from our optical and infrared spectra, which are poor in quality. $D_{\rm par}$=2409$^{+445}_{-329}$ pc agrees reasonably with $D_{\rm spec}$=1980 pc. $V_{\odot}$= $-$27 \kms\ from our infrared spectrum.

G012.3407$-$00.3949 (\#32) --- From the strong \ion{H}{1} lines and weak \ion{He}{1} and absent \ion{He}{2} in our infrared spectrum we classify this star (GID=4095629759211238400, $G$=13.8 mag) as early B.  The  \ion{He}{1} features appear doubled, with components at -126 \kms\ and 308 \kms, but the SNR is low and the velocities are considerably uncertain.   Our optical spectrum on 2018 June 13 looks double-featured in the \ion{He}{1} lines, with components at $-$133 \kms\ and 257 \kms.    The white-light guide camera images appear elongated east-west for this object, suggesting the possibility of a visual double with a sub-arcsecond separation.  We adopt B2V+B2V which provides good agreement between  $D_{\rm par}$=2398$^{+626}_{-419}$ pc and $D_{\rm spec}$=2560 pc.  $V_{\odot}$=32 \kms.  We have arbitrarily divided the mass loss rates for this object by two as a correction for the double nature of the source.   

G014.4703$-$00.6427 (\#46) --- (GID=4097416946633635584, $G$=16.5 mag) The EW of \ion{He}{1} 2.1126 $/mu$m is 1.01 \AA\  making this a probable  B2 star, likely a dwarf star given its parallax distance of  $D_{\rm par}$=688$^{+163}_{-111}$ pc.   With $H$=9.5 mag and $G$=16.5, the implied reddening is $A_{\rm K}$=1.2 mag ($\simeq$ $A_{\rm V}\simeq$11 mag.), unusually large for such a small distance.    $V_{\odot}$= $-$32 \kms, $-$17 \kms, and =$-$4 \kms\ on 2018 May 3, 2018 June 3, 2018 June 24, respectively from our infrared spectra, so we adopt an average of $-$18 \kms.

G017.0826+00.9744 (\#67) --- An O9V based on classification by \citet{Evans2005} (NGC6611ESL45; GID=4146617819936287488, $G$=11.5 mag) $D_{\rm par}$=1507$^{+123}_{-106}$ pc.  Our optical spectrum shows from 2018 June 13 yields $V_{\odot}$=+9 \kms\ and displays \ion{He}{2} 5410 with a EW of about 0.6 \AA, consistent with this classification and suggests $D_{\rm spec}$=2010 pc.  This target appears to be a visual double with a separation of about 1.5\arcsec, position angle PA$\simeq$50\degr, and a flux difference of at least five magnitudes and the brighter component to the northeast.   

G019.8107+00.0965 (\#83) --- (GID=4153376178622946560, $G$=14.3 mag) Our optical spectrum shows broad and very weak \ion{He}{1} lines, suggesting a mid--late B star.  A velocity measured from the broad (16 \AA\ FWHM) H$\alpha$ line yields $V_{\odot}$$\simeq$0 \kms. $D_{\rm par}$=2709$^{+318}_{-259}$ pc is consistent with a B2II at  $D_{\rm spec}$=3210 pc.   The luminosity class is very uncertain.

G023.0958+00.4411 (\#100) --- The optical spectrum of this star (GID=4156581976571298944, $G$=13.4 mag) is of poor quality, but the strength of \ion{He}{1} in a B2V yields good agreement with the spectrum and between $D_{\rm par}$=1819$^{+247}_{-195}$ pc and  $D_{\rm spec}$=1910 pc. $V_{\odot}$=$+$15 \kms.

G023.1100+00.5458 (\#101) --- B0V provides a good fit to the optical spectrum of this star (GID=4156594925877520512, $G$=12.1 mag). $D_{\rm par}$=2330$^{+460}_{-333}$ pc agrees well with  $D_{\rm spec}$=2043 pc. $V_{\odot}$= $+$14 \kms.

G027.3338+00.1784 (\#129) --- With a parallax distance of $D_{\rm par}$=870$^{+79}_{-66}$ pc this star (GID=4256681923166288640, $G$=10.1 mag) displays the CO bandhead features redward of 2.4 $\mu$m, characteristic of cool stars.  M0III provides a reasonable fit to  the optical spectral properties for a spectroscopic distance of $D_{\rm spec}$=930 pc. Cross correlating the K-band spectrum with a PHOENIX model atmosphere for $T=$4000~K, log($g$)=2.0, and solar metallicity results in a velocity of  $V_{\odot}$= $-$25 \kms. This is one of only three stars in the present sample that is not a hot, early-type star, although it may be a descendant of a massive star.  The central star of the nebula is bright and unambiguous, and there are no other $Gaia$ sources within 6\arcsec.   

G031.9308+00.2676 (\#163) --- We estimate this star (GID=4266110686393370112, $G$=13.2 mag) as B3IIe owing to the weak H$\alpha$ compared to the \ion{He}{1} lines in our optical spectrum.   The evolved luminosity class provides $D_{\rm spec}$=5760 pc, a somewhat questionable match to $D_{\rm par}$=4438$^{+849}_{-629}$ pc. This object exhibits a considerable redshift at $V_{\odot}$= $+$71 \kms. 

G037.2933+00.6703 (\#201) --- Our infrared spectrum of this star (GID=4281150081268339072, $G$=16.3 mag) shows  \ion{He}{1} 2.1126 $\mu$m in absorption.  The parallax distance of  $D_{\rm par}$=1088$^{+324}_{-205}$ pc agrees well with the spectrophotometric distance of $D_{\rm spec}$=1105 pc for a B2V.  Our optical spectrum of a slightly brighter $G$=15.1 star, GID=4281150081268339456, which lies about 3\arcsec\ to the west-southwest, is similar to a K3III, consistent with the parallax distance of 630 pc for that star. 

G055.5792+00.6749 (\#289) --- Our two infrared spectra of this star (GID=2017772886883963392, $G$=12.3 mag) from 2018 May 29 and 2018 June 03 show strong CO bandhead features near 2.4 $\mu$m. $V_{\odot}$=15 \kms\ and $V_{\odot}$=21 \kms\ on the two nights, respectively, by cross correlation with an M0III which provides a reasonable match to the spectral features and to the photometric properties ($D_{\rm spec}$=2150 pc and $D_{\rm par}$=2138$^{+256}_{-208}$ pc).  We adopt $V_{\odot}$=18 \kms.  The field contains multiple 24 $\mu$m and 70 $\mu$m arcs and prominent stars, creating considerable ambiguity about the geometry and nature of this object \citep[][, Figure 13.289]{Kobulnicky2016}.  

G073.6200+01.8522 (\#320) --- \Citet{Chick2019} gives B0III (HD191611; GID=2059236196250413696, $G$=8.5 mag) but O9III provides a better match between $D_{\rm par}$=2509$^{+254}_{-212}$ pc and $D_{\rm spec}$=2450 pc. $V_{\odot}$=3  \kms \citep{Chick2019}.

G074.3117+01.0041 (\#322) --- \citet{Chick2019} classifies this star (GID=2060507437839909248, $G$=12.6 mag ) as O8V, albeit with some uncertainty.  With $D_{\rm par}$=4460$^{+605}_{-481}$ pc, consistency with $D_{\rm spec}$=4462 pc is achieved by assigning O8IV, consistent with its spectrum.  $V_{\odot}$=$-$15 \kms\ \citep{Chick2019}. This object appears to lie inside of, but does not directly face an 8 $\mu$m bright-rimmed cloud.  

G077.0505$-$00.6094 (\#329) --- As an O8V (KGK2010-10; GID=2063868163830050176, $G$=14.1 mag)  $D_{\rm spec}$=2710 pc, slightly larger than the parallax distance  $D_{\rm par}$=2115$^{+280}_{-223}$ pc. $V_{\odot}$=$-$39 \kms\ \citep{Chick2019}.  A contribution from a hypothetical companion would easily resolve this discrepancy.

G078.2869+00.7780 (\#331) --- \citet{{Vijapurkar1993}} lists this star (LSII+39~53; GID=2067267299727875584, $G$=9.9 mag) as O7V which yields $D_{\rm spec}$=1760 pc, consistent with $D_{\rm par}$=1594$^{+116}_{-101}$ pc. $V_{\odot}$=$-$51 \kms\ \citet{Chick2019}.

G078.5197+01.0652 (\#333) ---   Our optical spectrum for this object (GID=2067382031192434816, $G$=13.0 mag) from 2018 June 13 shows doubled \ion{He}{1} features with approximately equal depth at $-$241 and +205 \kms.  There is no indication of \ion{He}{2}. It has $D_{\rm par}$=1554$^{+245}_{-188}$ pc, consistent with $D_{\rm spec}$=1410 pc for two B0V's.     

G079.8223+00.0959 (\#338) --- This probable O8V (CPR2002A10; GID=2064738049323468928, $G$=13.8 mag) has $D_{\rm par}$=1792$^{+225}_{-181}$ pc, consistent with $D_{\rm spec}$=1600 pc and  $V_{\odot}$=$-$10 \kms\ \citep{Chick2019}.

G080.2400+00.1354 (\#339) --- This O5V (CPR2002A37; GID=2064838375463800448, $G$=11.7 mag) has $D_{\rm par}$=1703$^{+129}_{-112}$ pc, consistent with $D_{\rm spec}$=1590 pc.   $V_{\odot}$=$-$43 \kms\ \citep{Chick2019}.

G080.8621+00.9749 (\#341) --- Classified as O9V by \citet{Chick2019} this star (KGK2010-1; GID=2067963015711183744, $G$=12.1 mag) has $D_{\rm par}$=1622$^{+82}_{-73}$ pc, consistent with $D_{\rm spec}$=1590 pc. $V_{\odot}$$-$38 \kms\ \citet{Chick2019}.

G080.9020+00.9828 (\#342) --- This mid-B star (KGK2010-2; GID=2067963535403817728, $G$=14.5 mag) displays weak \ion{He}{1} in our optical and infrared spectra, but strong \ion{H}{1} features. B4V fits the spectral features and provides consistency with  between the parallax distance of $D_{\rm par}$=1538$^{+82}_{-73}$ pc   and the $D_{\rm spec}$=1720 pc.   Our infrared spectrum of this star on 2018 June 03 indicates  $V_{\odot}$=$-$17 \kms\ which compares favorably to  the average value of optical spectra from \citet{Chick2019}, $V_{\odot}$=$-$7 \kms .

G082.4100+02.3254 (\#344) --- This well-studied O4If (BD+43 3654; GID=2069819545390584192, $G$=9.1  mag) has $D_{\rm par}$=1577$^{+81}_{-74}$ pc, consistent with $D_{\rm spec}$=1460 pc.  \citet{Chick2019} reports $V_{\odot}$=$-$36 \kms.

G104.3447+02.2299 (\#353) --- A strong \ion{He}{2} 5410 \AA\ makes this star (GID=2201205412482296448, $G$=11.6 mag) a probable O5V, yielding consistency between $D_{\rm par}$=3932$^{+468}_{-380}$ pc and $D_{\rm spec}$=4370 pc.  $V_{\odot}$=$-$47 \kms\ from \citet{Chick2019}. H$\alpha$ appears in emission and possibly variable.  

G106.6327+00.3917 (\#356) --- An early B temperature class (HD240015; GID=2008430237808841600, $G$=9.8 mag), the parallax distance of $D_{\rm par}$=2583$^{+226}_{-193}$ pc requires  about B0III  for consistency at D$_{spec}$=2690 pc. $V_{\odot}$=$-$54 \kms \citep{Chick2019}. 

G106.6375+00.3783 (\#357) --- (HD240016; GID=2008383302396655488; G=9.4 mag) $D_{\rm par}$=1493$^{+72}_{-65}$ pc is  consistent with $D_{\rm spec}$=1410 pc for a B2III.  $V_{\odot}$=$-$88 \kms \citet{Chick2019}.

G108.9891+01.5606 (\#361) ---  A probable B1I,  this star (TYC 4278-522-1; GID=2206818556775783552, $G$=10.1 mag) displays H$\alpha$ in emission and has $D_{\rm par}$=3896$^{+515}_{-412}$ pc, roughly consistent with $D_{\rm spec}$=4870 pc given the large uncertainties inherent in assigning magnitudes to supergiants.  $V_{\odot}$=$-$53 \kms \citep{Chick2019}.

G109.1157+00.6799 (\#362) --- This very distant ($D_{\rm par}$=5295$^{+944}_{-714}$ pc) object (GID=2014562519088745344, $G$=12.7 mag) is a probable B2I or similar. $V_{\odot}$=$-$33 \kms \citet{Chick2019}.

G133.1567+00.0432 (\#367) --- Strong \ion{He}{2} matches an O7V at $D_{\rm spec}$=2050 pc for this star (LSI+60 226; GID=507686819685070208, $G$=10.4 mag) consistent with $D_{\rm par}$=2131$^{+130}_{-116}$ pc.    Our two infrared spectra on 2017 July 13 and 2017 Sep 01 yield $V_{\odot}$=$-$61 \kms\ and  $V_{\odot}$=46 \kms, suggesting that this is a single-lined spectroscopic binary.

G134.3552+00.8182 (\#368) --- Classified as O9.5V by \citet{Massey1995}, this eruptive variable star (KM Cas; GID=465523778576137600, $G$=10.8 mag) appears somewhat hotter in our 2018 June 13 optical spectrum, perhaps O8.5V.  H$\alpha$ is very weak, suggesting emission has filled some \ion{H}{1} features.  It has $D_{\rm par}$=2298$^{+187}_{-161}$ pc, somewhat larger that the spectrophotometric distance of $D_{\rm spec}$=1690 pc for an O9.5V but consistent with the  $D_{\rm spec}$=2040 pc for an O8.5V, or it may be slightly evolved and more luminous as an O8.5IV.  $V_{\odot}$=$-$1 \kms\ from our infrared and optical spectra.  There are additional absorption features present across the spectrum that are inconsistent with models of hot stars and may indicate circumstellar material.  

G137.4203+01.2792 (\#369) --- \citet{Conti1974} list this star (BD+60 586; GID=464697873547937664, $G$=8.4 mag) as O8III, which leads to a $D_{\rm spec}$=2790 pc, in better agreement with the $D_{\rm par}$=2762$^{+427}_{-329}$ pc than the O7.5V given by \citet{Hillwig2006}. O8III provides an excellent match to our optical spectrum of 2018 June 13.  We adopt $V_{\odot}$=$-$40 \kms\ from our optical spectrum, consistent with the $V_{\odot}$=$-$49 \kms\ from the literature.

G223.7092$-$01.9008 (\#380) --- We adopt the B0IVe (HD53367; GID=3046530911350220416, $G$=7.0 mag) given by \citet{Tjin2001} which predicts  $D_{\rm spec}$=369 pc, considerably greater than the $D_{\rm par}$=129$^{+12}_{-14}$ pc, however $Gaia$ parallaxes for very bright stars are especially uncertain as of this writing. We also adopt $V_{\odot}$=21 \kms\ from the literature.  A B1V would be required in order to reconcile the parallax and spectrophotometric distances, so we provisionally adopt this classification.  

G224.1685$-$00.7784 (\#381) ---  We adopt the O7Vz given by \citet{Sota2014} for this star (HD54662; GID=3046582725837564800, $G$=6.1 mag).  This yields $D_{\rm spec}$=985 pc, in good agreement with $D_{\rm par}$=1142$^{+120}_{-100}$ pc.  We also adopt $V_{\odot}$=57 \kms\ from the literature. The possible double-lined nature (O6.5V+O7V--O9V 2119 d period)  of this source reported by \citet{Boyajian2007} was not confirmed by \cite{Sota2014}.

G224.7096$-$01.7938(\#382)  ---  The parallax and proper motion for this  B0III \citep{Tjin2001} star (FN CMa; GID=3046209987096803584, $G$=5.3 mag) comes from the HIPPARCOS mission. The parallax distance of $D_{\rm par}$=934 pc agrees well with the spectrophotometric distance of 637 pc.  We adopt $V_{\odot}$=31 \kms\ from the literature.

G253.7104$-$00.1920 (\#384) --- We adopt O7IIInn from \citet{Vijapurkar1993} for this star (CD-35 4415 ; GID=5543183649492723584, $G$=10.3 mag) $D_{\rm par}$=3791$^{+463}_{-375}$ pc, consistent with $D_{\rm spec}$=3580 pc.  

G268.9550$-$01.9022 (\#391) ---  We adopt B7Iab from \citet{Houk1978} for this Algol-type eclipsing binary (HD77207 ; GID=5325452481440825600, $G$=9.2 mag). $D_{\rm par}$=1851$^{+106}_{-95}$ pc, vastly smaller than the $D_{\rm spec}$=6900 pc. A few such large discrepancies are not surprising given the inherent difficulty in assigning luminosities to supergiants.  

G269.2089$-$00.9138 (\#392) --- Listed generically in the literature as an OB star (CPD-47 3051; GID=5326951150144920064,  $G$=10.8 mag), the parallax distance of $D_{\rm par}$=1901$^{+121}_{-136}$ pc and photometric data are consistent with a B0V ($D_{\rm spec}$=2200 pc).  

G272.5794$-$01.7247 (\#394) --- We adopt B0 from \citet{Reed2003} for this object (HD298310; GID=5313479796251809664, $G$=9.8 mag). A B0III provides $D_{\rm spec}$=2010 pc, consistent with    $D_{\rm par}$=2199$^{+155}_{-136}$ pc.  

G273.1192$-$01.9620 (\#395) --- \citet{Reed2003} lists this (HD298353; GID=5311906841793959936, $G$=9.8 mag) as an O star.   We adopt O7V which provides $D_{\rm spec}$=2450 pc, consistent with   $D_{\rm par}$=2685$^{+296}_{-244}$ pc.  

G282.1647$-$00.0256 (\#397) --- For this star (CD-55 3196; GID=5259053730157539328, $G$=10.5) we adopt O9.5III \citep{Parthasarathy2012} which produces good agreement between $D_{\rm par}$=2513$^{+186}_{-163}$ pc and $D_{\rm spec}$=2320 pc. 

G286.0588$-$01.6633 (\#404) --- This star (CD-59 3123; GID=5255179905867530752, $G$=9.1 mag) is listed as O9.5Ib in \citet{Vijapurkar1993}.   With a parallax distance of $D_{\rm par}$=3969$^{+470}_{-545}$ pc  its photometric data would be consistent with the spectral type and luminosity class, $D_{\rm spec}$=3601 pc.  The stellar driver of this bowshock was incorrectly identified as Tyc8613-707-1 in \citet[][Figure~13.404; the brighter source to the upper right of Tyc8613-707-1 at center is CD-59 3123]{Kobulnicky2016} and \citet{Kobulnicky2017}.  The position and and identification is corrected in this work.

G286.4644$-$00.3478 (\#405) --- Listed as B5 in \citet{Nesterov1995} this star (HD303197; GID=5350640192585837568, $G$=9.1 mag) has $D_{\rm par}$=2328$^{+154}_{-136}$ pc consistent with a B3III at that distance  ($D_{\rm spec}$=2270 pc) but inconsistent with any early B dwarf.  

G287.1148$-$01.0236 (\#406) --- HD92607 (GID=5254478593582508288, $G$=8.1 mag) is an O9Vn \citep{Sota2014}, predicting a $D_{\rm spec}$=2310 pc, broadly consistent with $D_{\rm par}$=2786$^{+597}_{-424}$ pc. 

G287.4071$-$00.3593 (\#407) --- HD93249 (GID=5350395383778733568, $G$=8.3 mag), an O9III+O binary  \citep{Sota2014}, has  $D_{\rm par}$=3096$^{+376}_{-304}$ pc, consistent with the expected $D_{\rm spec}$=2880 pc.

G287.6131$-$01.1302 (\#409) --- HD93027 (GID=5254268518156437888, $G$=8.7 mag), an O9.5IV \citep{Sota2011} has $D_{\rm spec}$=2940 pc, similar to the $D_{\rm par}$=3426$^{+507}_{-395}$ pc. This is a single-lined eclipsing binary according to \citet{Sota2011}, suggesting a significant difference in mass between the primary and secondary star.

G287.6736$-$01.0093 (\#410) --- HD305536 (GID=5254269961265754368, $G$=9.0 mag) is an O9.5V+unknown  binary candidate\citep{Sota2014, Levato1990}.   $D_{\rm par}$=2266$^{+152}_{-134}$ pc is consistent with $D_{\rm spec}$=2120 pc. 

G288.1505$-$00.5059  (\#411)--- HD305599 (GID=5338310887667141888, $G$=9.9 mag) lists as O9.5V \citep{Alexander2016}.  $D_{\rm par}$=2071$^{+145}_{-127}$ pc agrees roughly with  $D_{\rm spec}$=2500 pc.  

G288.3138$-$01.3085 (\#413)  --- HD93683 (GID=5242187050013731200, $G$=7.8 mag) is given as an O9V+B0V binary \citep{Alexander2016}.  $D_{\rm par}$=1629$^{+248}_{-191}$ pc is somewhat larger than  $D_{\rm spec}$=1090 pc, even accounting for the possible binary, raising the possibility that one or both components may be slightly evolved.  O9IV+B0IV would provide a closer match to the parallax distance and the observed mass-loss rate.  

G288.4263$-$01.2245 (\#414) --- HD93858 (GID=5242184954087251456, $G$=9.0 mag) is an eclipsing binary with a B3II/III primary \citep{Houk1978}.    $D_{\rm par}$=2602$^{+196}_{-191}$ pc, broadly consistent with this luminosity class.  

G308.0406+00.2473 (\#463) --- TYC8995-1548-1 (GID=5865357598905884288, $G$=11.6 mag) lists as an OB star \citep{Stephenson1971} and B0III provides a good match to $D_{\rm par}$=3689$^{+618}_{-470}$ pc. 

G326.7256+00.7742 (\#555) --- ALS 18049 (GID=5885668499206748160, $G$=12.2 mag) lists in the literature as O7V, but O9V provides a better match between  $D_{\rm spec}$=3120 pc and $D_{\rm par}$=2677$^{+421}_{-323}$ pc.  

G332.4863+00.8256 (\#589) --- CD-49 10393 (GID=5935169268617561344, $G$=11.1 mag) lists as an OB star \citep{Stephenson1971}.  B0V provides good agreement with $D_{\rm par}$=2533$^{+306}_{-348}$ pc.  

G340.4772$-$00.1528 (\#624) --- This star's (GID=5964264231754517504, $G$=13.3 mag) infrared spectrum display Br $\gamma$ in emission, but \ion{He}{1} 2.1126 $\mu$m in  absorption and no \ion{He}{2} making it a probable B emission line star. $D_{\rm par}$=3508$^{+675}_{-496}$ pc which requires  something close to a B0IIIe for a consistent spectrophotometric distance. $V_{\odot}$=+10 \kms. 

G340.8579$-$00.8793 (\#626) --- This object (GID=5964030074446095744, $G$=11.3 mag) display the Bracket series in absorption, \ion{He}{1} 2.1126 $\mu$min absorption, and weak \ion{He}{2} 2.1885 $\mu$m, suggesting a late O star.  $D_{\rm par}$=2169$^{+389}_{-288}$ pc which is close to an O9V at $D_{\rm spec}$=2290 pc.     $V_{\odot}$=$-$13 \kms.

G342.3422$-$00.4456 (\#634) --- (HD152756; GID=5964575672745370752, $G$=8.8 mag) shows strong \ion{He}{1} and Bracket series absorption  but no \ion{He}{2} is present in the K-band spectrum. $D_{\rm par}$=1802$^{+220}_{-178}$ pc which is an excellent match to a B0III at $D_{\rm spec}$=1800 pc.   $V_{\odot}$=$-$11 \kms.

G342.5873+00.1600 (\#635) --- A probable late O star (GID=5964883879575434240, $G$=13.5 mag) showing \ion{He}{2} with an EW of 1.2 \AA, an O6V provides a good match to the parallax distance of  $D_{\rm par}$=2809$^{+829}_{-544}$ pc and to the spectrum. $V_{\odot}$=$-$26 \kms.

G342.7172$-$00.4361 (\#637) --- Displaying \ion{He}{1} lines, this star (GID=5964678060440055296, $G$=15.2 mag) is nearby at  $D_{\rm par}$=1315$^{+193}_{-150}$ pc 1346.  The infrared spectrum and distance agrees well with a B2V at $D_{\rm spec}$=1260 pc. $V_{\odot}$=$-$38 \kms.

G344.4658$-$00.5580 (\#648) --- (GID=5965562308305643008; $G$=14.3 mag)  This star shows H$\alpha$ in emission with no detectable helium lines, making the spectral type particularly uncertain.  We estimate it near B0III, primarily on the basis of $D_{\rm spec}$=1877 pc agreeing with $D_{\rm par}$=1893$^{+597}_{-373}$ pc. 
 
G346.1388$-$00.2184 (\#653) --- This target (GID=5966132787371809536; $G$=13.7; $H$=10.2 mag) has weak narrow H$\alpha$ and  weak features near 4383/85 \AA\ in our optical spectrum that are probable iron lines. The spectral features suggest that this is a late-G or early-K spectral type, making it unlikely to be the star driving a bowshock nebula  at $D_{\rm par}$=1664$^{+114}_{-100}$ pc.  Such a distance would require a giant (K0III:) luminosity class for consistency with the optical and IR magnitudes.    Five other $Gaia$ sources within 8\arcsec\ are 3--7  magnitudes fainter. There is no other bright early type candidate driving star along symmetry axis.     The $G$=19.4 ($H$=13.8) star GID=5966132787358462848 is a candidate central star located 6\arcsec\ to the N-NE.   Given the lack of an identifiable early type driving star of an appropriate magnitude at this distance, the nature of this arcuate nebula is uncertain. 

G346.2958+00.0744 (\#655) ---  This probable B1V star (GID=5966890930670155520; $G$=13.7 mag)  has $D_{\rm spec}$=2650 pc, consistent with the $D_{\rm par}$=2168$^{+251}_{-204}$ pc.  The \ion{He}{1} lines in our optical spectrum appear asymmetric, so this may be a double-lined binary with blended components.   It has $V_{\odot}$= $-$69 \kms.  

G349.5431$-$00.5952 (\#673) --- This B1V (GID=5972791773374952192; $G$=13.5 mag) has $D_{\rm spec}$=2900 pc, in excellent agreement with  $D_{\rm par}$=2705$^{+337}_{-272}$ pc.   $V_{\odot}$= $-$22 \kms\ from the average of our optical and infrared radial velocities.  

G353.4162+00.4482 (\#692) ---  Our optical spectrum of this star (GID=5975945894262363776; $G$=10.7 mag) is well fit by an O7V, in good agreement with \citet[][; O7.5V]{Gvaramadze2011}.  $D_{\rm spec}$=1810 pc consistent with $D_{\rm par}$=1717$^{+171}_{-142}$.   The He lines are slightly asymmetric suggesting a possible blended binary.   $V_{\odot}$= $-$30 \kms\ assuming a single line profile.

G355.4972$-$00.7571 (\#694) --- (GID=4053776191953892224; $G$=12.1 mag) B2V matches the narrow He lines and hint of \ion{Mg}{2} 4481 \AA\ in our optical spectrum.  $D_{\rm spec}$= 1570 pc, consistent with $D_{\rm par}$=1329$^{+233}_{-173}$ pc.   $V_{\odot}$= $-$11.2 \kms.

G356.6602+00.9209 (\#700) --- This B0V (GID=4055297370601329024; $G$=13.1 mag) from our optical spectrum has $D_{\rm spec}$=3200 in good  agreement with $D_{\rm par}$=3370$^{+664}_{-481}$ pc.   $V_{\odot}$=$-$25.2 \kms.  

G359.9536$-$00.5088 (\#709) --- This source  (GID=4057277728502494080; $G$=11.4 mag) displays doubled \ion{He}{1} 5876 \AA\ lines in our optical spectrum in a ratio of about 2:3 at  $V_{\odot}$=-133 and 115 \kms, respectively.  $D_{\rm spec}$=2860 for a B0V, in reasonable agreement with $D_{\rm par}$=3091$^{+539}_{-403}$ pc, given the probable binary nature.

\acknowledgments This work has been supported by the National Science Foundation through grants AST-1412845 and AST-1411851. We offer our abundant gratitude to the reviewer, Jo Puls, for several detailed readings and many expert suggestions that substantially improved this work.  Curt Struck alerted us to the work of William Henney on radiation-driven bowshocks.  Exchanges with William Henney contributed greatly to this manuscript and prevented an error in the selection of dust grain emission coefficients.  We also thank Rico Ignace, Derek Massa, and Eigen Li for their input at various stages of completion.    

\vspace{5mm}
\facilities{APO, WIRO, SST, HSO, $Gaia$}
\software{IRAF, {\citep{Tody1986}},
 SpeXtool, {\citep{Cushing2004}}
 }
 
\newpage

\begin{longrotatetable}
\begin{deluxetable}{rcrcrrrrrrrrrlr}
\tablecaption{Measured \& adopted parameters for stars and their bowshock nebulae \label{tab:basic}}
\tabletypesize\scriptsize 
\tablehead{
\colhead{ID} &\colhead{Name}&\colhead{Alt. name}&\colhead{Sp.T.}&\colhead{T$_{eff}$}&\colhead{R$_*$ }	 &\colhead{Mass}&\colhead{$V_\infty$}  &\colhead{D}  &\colhead{$R_0$}	&\colhead{$R_0$}&\colhead{Peak$_{70}$}  	&\colhead{$\ell$} & \colhead{obs.} & \colhead{$L_*/L_{\rm IR}$}   \\
\colhead{}   &\colhead{    }&\colhead{         }&\colhead{ }   &\colhead{(K)}	   &\colhead{($R_\odot$)}&\colhead{($M_\odot$)}    &\colhead{(\kms)}       &\colhead{(pc)} &\colhead{(arcsec)} &\colhead{(pc)} &\colhead{(10$^7$ Jy sr$^{-1}$)}&\colhead{(arcsec)} &\colhead{}  &\colhead{} \\
\colhead{(1)}&\colhead{(2)} &\colhead{(3)}	&\colhead{(4)}  &\colhead{(5)} &\colhead{(6)}	   &\colhead{(7)}	 &\colhead{(8)}       &\colhead{(9)}	      &\colhead{(10)}  &\colhead{(11)}&\colhead{(12)}	 &\colhead{(13)}  &\colhead{(14)}  &\colhead{(15)} }   
\startdata
1 & \nodata & G000.1169$-$00.5703 & O8V & 35500 & 9.4 & 26.0 & 2000.0 & 2441 & 26.4 & 0.344 & 166 & 39 & 0.3 & 252 \\
3 & \nodata & G001.0563$-$00.1499 & B5III: & 15000 & 8.0 & 7.0 & 500.0 & 1527 & 6.2 & 0.05 & 432 & 12 & 0.4 & 11 \\
7 & \nodata & G003.8417$-$01.0440 & B2V & 20600 & 5.4 & 10.9 & 700.0 & 1313 & 8.6 & 0.06 & 764 & 27 & 0.4 & 61 \\
11 & \nodata & G005.5941+00.7335 & B0V & 29000 & 7.4 & 17.5 & 1200.0 & 2446 & 7.6 & 0.099 & 15 & 13 & 0.3 & 1389 \\
13 & Zeta Oph & G006.2812+23.5877 & O9.2IV & 31000 & 7.2 & 19.0 & 1300.0 & 112 & 299.0 & 0.179 & 12 & 277 & 0.3 & 1735 \\
16 & \nodata & G006.8933+00.0743 & B0III & 29000 & 15.0 & 20.0 & 1200.0 & 3008 & 29.0 & 0.465 & 83 & 23 & 0.3 & 807 \\
26 & \nodata & G011.0709$-$00.5437 & O9V & 31500 & 7.7 & 18.0 & 1200.0 & 3060 & 8.2 & 0.134 & 24 & 7 & 0.4 & 532 \\
28 & \nodata & G011.6548+00.4943 & B0V & 29000 & 7.4 & 17.5 & 1200.0 & 2409 & 9.2 & 0.118 & 33 & 26 & 0.3 & 330 \\
32 & \nodata & G012.3407$-$00.3949 & B2V+B2V & 20600 & 5.4 & 10.9 & 800.0 & 2398 & 7.5 & 0.096 & 64 & 13 & 0.4 & 73 \\
46 & \nodata & G014.4703$-$00.6427 & B2V: & 20600 & 5.4 & 10.9 & 700.0 & 688 & 14.7 & 0.054 & 33 & 28 & 0.4 & 45 \\
67 & NGC6611ESL45 & G017.0826+00.9744 & O9V & 31500 & 7.7 & 18.0 & 1300.0 & 1507 & 7.5 & 0.06 & 256 & 7 & 0.3 & 44 \\
83 & \nodata & G019.8107+00.0965 & B8II & 11000 & 15.0 & 20.0 & 700.0 & 2709 & 5.6 & 0.081 & 49 & 11 & 0.5 & 460 \\
100 & \nodata & G023.0958+00.4411 & B1V & 26000 & 6.4 & 14.2 & 800.0 & 1819 & 6.7 & 0.065 & 39 & 12 & 0.3 & 144 \\
101 & \nodata & G023.1100+00.5458 & B0V & 29000 & 7.4 & 17.5 & 1200.0 & 2330 & 12.7 & 0.158 & 199 & 34 & 0.3 & 48 \\
129 & TYC5121-625-1 & G027.3338+00.1784 & M0III & 3800 & 40.0 & 1.2 & \nodata.0 & 870 & 11.2 & 0.052 & 232 & 11 & 0.0 & 4 \\
163 & \nodata & G031.9308+00.2676 & B3II & 17000 & 11.0 & 14.0 & 600.0 & 4438 & 15.5 & 0.367 & 33 & 18 & 0.5 & 31 \\
201 & \nodata & G037.2933+00.6703 & B2V & 20600 & 5.4 & 10.9 & 800.0 & 1088 & 7.6 & 0.044 & 36 & 13 & 0.0 & 162 \\
289 & \nodata & G055.5792+00.6749 & M0III & 3800 & 40.0 & 1.2 & \nodata.0 & 2138 & 34.0 & 0.388 & 38 & 33 & 0.0 & 0 \\
320 & HD191611 & G073.6200+01.8522 & O9III & 30700 & 13.6 & 23.0 & 1300.0 & 2509 & 35.5 & 0.475 & 53 & 21 & 0.4 & 1056 \\
322 & \nodata & G074.3117+01.0041 & O8IV: & 32500 & 14.1 & 27.0 & 2000.0 & 4460 & 19.5 & 0.464 & 46 & 20 & 0.3 & 1998 \\
% 329 & KGK2010-10 & G077.0505$-$00.6094 & O8V & 33400 & 8.5 & 22.0 & 2000.0 & 2115 & 10.0 & 0.113 & 61 & 27 & 0.3 & 1357 \\
331 & LSII+39~53 & G078.2869+00.7780 & O7V & 35500 & 9.3 & 26.0 & 2500.0 & 1594 & 42.0 & 0.357 & 48 & 55 & 0.3 & 296 \\
333 & \nodata & G078.5197+01.0652 & B0V+B0V & 30900 & 7.4 & 17.5 & 2500.0 & 1554 & 28.9 & 0.24 & 66 & 16 & 0.4 & 461 \\
338 & CPR2002A10 & G079.8223+00.0959 & O8V: & 33400 & 8.5 & 22.0 & 2000.0 & 1792 & 23.0 & 0.22 & 319 & 29 & 0.3 & 138 \\
339 & CPR2002A37 & G080.2400+00.1354 & O5V & 41500 & 11.1 & 37.0 & 2900.0 & 1703 & 70.0 & 0.636 & 103 & 47 & 0.3 & 640 \\
341 & KGK2010-1 & G080.8621+00.9749 & O9V & 31500 & 7.7 & 18.0 & 1300.0 & 1622 & 20.0 & 0.173 & 19 & 31 & 0.4 & 379 \\
342 & KGK2010-2 & G080.9020+00.9828 & B4V & 15500 & 4.3 & 6.5 & 800.0 & 1538 & 10.0 & 0.082 & 231 & 14 & 0.4 & 22 \\
344 & BD+43~3654 & G082.4100+02.3254 & O4If & 40700 & 19.0 & 58.0 & 3000.0 & 1577 & 193.0 & 1.623 & 234 & 170 & 0.3 & 247 \\
353 & \nodata & G104.3447+02.2299 & O5V & 41500 & 11.1 & 37.0 & 2900.0 & 3932 & 30.5 & 0.64 & 19 & 20 & 0.4 & 8563 \\
356 & HD240015 & G106.6327+00.3917 & B0III & 29000 & 15.0 & 20.0 & 1000.0 & 2583 & 15.6 & 0.215 & 13 & 15 & 0.4 & 22422 \\
357 & HD240016 & G106.6375+00.3783 & B2III & 20900 & 15.0 & 20.0 & 700.0 & 1493 & 21.8 & 0.174 & 13 & 25 & 0.3 & 16066 \\
361 & TYC4278-522-1 & G108.9891+01.5606 & B1Ia & 20000 & 25.0 & 25.0 & 800.0 & 3896 & 91.7 & 1.905 & 13 & 75 & 0.5 & 1974 \\
362 & \nodata & G109.1157+00.6799 & B2I & 17600 & 40.0 & 23.0 & 700.0 & 5295 & 14.7 & 0.415 & 16 & 27 & 0.4 & 1999 \\
367 & LSI+60~226 & G133.1567+00.0432 & O7V & 35500 & 9.4 & 26.0 & 2500.0 & 2131 & 36.0 & 0.409 & 28 & 51 & 0.3 & 4812 \\
368 & KM~Cass & G134.3552+00.8182 & O8.5V & 32500 & 8.5 & 22.0 & 2000.0 & 2298 & 14.0 & 0.172 & 99 & 22 & 0.3 & 3804 \\
369 & BD+60~586 & G137.4203+01.2792 & O8III & 32500 & 14.0 & 27.0 & 2000.0 & 2762 & 73.0 & 1.075 & 31 & 39 & 0.3 & 646 \\
380 & HD53367 & G223.7092$-$01.9008 & B2Ve: & 20600 & 5.4 & 17.5 & 700.0 & 130 & 22.0 & 0.015 & 143 & 49 & 0.3 & 2479 \\
381 & HD54662 & G224.1685$-$00.7784 & O7Vz & 35500 & 9.4 & 26.0 & 2500.0 & 1142 & 220.0 & 1.34 & 18 & 200 & 0.4 & 688 \\
382 & FN~Cma & G224.7096$-$01.7938 & B0III & 28000 & 15.0 & 20.0 & 1200.0 & 934 & 101.0 & 0.503 & 44 & 70 & 0.8 & 1547 \\
384 & CD-35~4415 & G253.7104$-$00.1920 & O7IIInn & 34600 & 14.5 & 31.0 & 2500.0 & 3791 & 31.0 & 0.627 & 93 & 23 & 0.3 & 1204 \\
391 & HD77207 & G268.9550$-$01.9022 & B7Iab + & 11000 & 50.0 & 20.0 & 500.0 & 1851 & 32.0 & 0.316 & 16 & 27 & 0.5 & 1863 \\
392 & CPD-47~3051 & G269.2089$-$00.9138 & B0V & 29000 & 7.4 & 17.5 & 1200.0 & 1901 & 28.0 & 0.284 & 21 & 39 & 0.3 & 691 \\
394 & HD298310 & G272.5794$-$01.7247 & B0V & 29000 & 15.0 & 20.0 & 1200.0 & 2199 & 21.0 & 0.246 & 26 & 45 & 0.4 & 867 \\
395 & HD298353 & G273.1192$-$01.9620 & O7V & 35500 & 9.4 & 26.0 & 2500.0 & 2685 & 32.0 & 0.458 & 33 & 23 & 0.3 & 1794 \\
397 & CD-55~3196 & G282.1647$-$00.0256 & O9.5III & 30200 & 13.4 & 21.0 & 1200.0 & 2513 & 22.0 & 0.295 & 99 & 28 & 0.3 & 288 \\
404 & CD-59~3123 & G286.0498$-$01.6583 & O9.5Ib & 28400 & 23.0 & 30.0 & 1200.0 & 3969 & 54.0 & 1.143 & 31 & 70 & 0.4 & 449 \\
405 & HD303197 & G286.4644$-$00.3478 & B3III & 17000 & 11.0 & 14.0 & 600.0 & 2328 & 21.0 & 0.261 & 49 & 27 & 0.4 & 44 \\
406 & HD92607 & G287.1148$-$01.0236 & O9IVn & 31100 & 10.0 & 20.0 & 1300.0 & 2786 & 15.8 & 0.235 & 116 & 26 & 0.3 & 195 \\
407 & HD93249 & G287.4071$-$00.3593 & O9III+O & 30700 & 13.6 & 23.0 & 1300.0 & 3096 & 7.8 & 0.129 & 232 & 25 & 0.3 & 88 \\
409 & HD93027 & G287.6131$-$01.1302 & O9.5IV & 30300 & 7.4 & 16.0 & 1200.0 & 3426 & 7.4 & 0.135 & 83 & 17 & 0.3 & 147 \\
410 & HD305536 & G287.6736$-$01.0093 & O9.5V+ B & 30500 & 7.4 & 16.0 & 1200.0 & 2266 & 3.7 & 0.045 & 366 & 14 & 0.3 & 227 \\
411 & HD305599 & G288.1505$-$00.5059 & O9.5V & 30500 & 7.4 & 16.0 & 1200.0 & 2071 & 4.2 & 0.046 & 166 & 6 & 0.3 & 757 \\
413 & HD93683 & G288.3138$-$01.3085 & O9V+B0V & 31500 & 7.7 & 18.0 & 1300.0 & 1629 & 15.0 & 0.13 & 61 & 28 & 0.3 & 709 \\
414 & HD93858 & G288.4263$-$01.2245 & B3II & 17000 & 11.0 & 14.0 & 600.0 & 2602 & 5.8 & 0.08 & 26 & 13 & 0.4 & 1481 \\
463 & TYC8995-1548-1 & G308.0406+00.2473 & B0III: & 28000 & 15.0 & 20.0 & 1200.0 & 3689 & 15.9 & 0.313 & 26 & 19 & 0.4 & 2459 \\
555 & ALS18049 & G326.7256+00.7742 & O9V: & 31500 & 7.7 & 18.0 & 1300.0 & 2677 & 4.1 & 0.059 & 465 & 14 & 0.3 & 16 \\
589 & CD-49 10393 & G332.4863+00.8256 & B0V & 29000 & 7.4 & 17.5 & 1200.0 & 2533 & 9.5 & 0.128 & 51 & 37 & 0.4 & 219 \\
624 & \nodata & G340.4772$-$00.1528 & B0III & 29000 & 15.0 & 20.0 & 1200.0 & 3508 & 18.4 & 0.344 & 49 & 38 & 0.3 & 515 \\
626 & \nodata & G340.8579$-$00.8793 & O9V & 31500 & 7.7 & 18.0 & 1300.0 & 2169 & 11.5 & 0.133 & 698 & 24 & 0.3 & 20 \\
634 & HD152756 & G342.3422$-$00.4456 & B0III & 28000 & 15.0 & 20.0 & 1200.0 & 1802 & 32.0 & 0.308 & 33 & 43 & 0.4 & 309 \\
635 & \nodata & G342.5873+00.1600 & O6V & 38100 & 10.2 & 32.0 & 2800.0 & 3197 & 31.0 & 0.529 & 96 & 45 & 0.5 & 83 \\
637 & \nodata & G342.7172$-$00.4361 & B2V & 20600 & 5.4 & 10.9 & 700.0 & 1315 & 5.8 & 0.041 & 8 & 7 & 0.3 & 210 \\
648 & \nodata & G344.4658$-$00.5580 & B0III & 29000 & 15.0 & 20.0 & 1200.0 & 1893 & 14.8 & 0.149 & 68 & 44 & 0.3 & 530 \\
653 & \nodata & G346.1388$-$00.2184 & K0III: & 4000 & 15.0 & 1.1 & \nodata.0 & 1664 & 1.9 & 0.017 & 31 & 14 & 0.0 & 1 \\
655 & \nodata & G346.2958+00.0744 & B1V: & 26000 & 6.4 & 14.2 & 800.0 & 2168 & 3.5 & 0.04 & 66 & 13 & 0.3 & 366 \\
673 & \nodata & G349.5431$-$00.5952 & B1V: & 26000 & 6.4 & 14.2 & 800.0 & 2705 & 6.0 & 0.087 & 23 & 15 & 0.3 & 1200 \\
692 & \nodata & G353.4162+00.4482 & O7V & 35500 & 9.3 & 26.0 & 2500.0 & 1717 & 34.0 & 0.311 & 515 & 55 & 0.4 & 102 \\
694 & \nodata & G355.4972$-$00.7571 & B2V: & 20900 & 5.4 & 10.9 & 700.0 & 1329 & 5.8 & 0.041 & 21 & 14 & 0.3 & 704 \\
700 & \nodata & G356.6602+00.9209 & B0V & 29000 & 7.4 & 17.5 & 1200.0 & 3370 & 15.8 & 0.284 & 13 & 21 & 0.3 & 1202 \\
709 & \nodata & G359.9536$-$00.5088 & B0V+B1V & 29000 & 7.4 & 17.5 & 1200.0 & 3091 & 11.2 & 0.185 & 16 & 11 & 0.4 & 49 
\enddata
 \tablecomments{(1) Identifier from \citet{Kobulnicky2016}, (2) Common name, (3) generic identifier in Galactic coordinates, 
 (4) spectral classification, primarily from this work and citations herein (5) adopted effective temperature based on spectral classification 
 using the theoretical scale of \citet{Martins2005a} and extrapolated to early B stars as in \citet{Pecaut2013}, (6) stellar radius based on spectral classification using the theoretical scale of \citet{Martins2005a}, (7) adopted stellar mass, (8) adopted terminal wind velocity from \citet{Mokiem2007}, (9) adopted distance, from $Gaia$ parallax-based distances estimated by \citet{Bailer-Jones2018}  (10) standoff distance in arcsec, (11) standoff distance in pc using the adopted distance and angular size adjusted by a statistical factor of 1.1 for projection effects, (12) peak 70 $\mu$m surface 
 brightness above adjacent background, (13) angular diameter of the nebula in arcsec defined by a chord intersecting the location of 
 peak surface brightness, (14) source of spectroscopy, described in the text, (15) ratio of stellar to infrared nebular luminosity.
 }
\tablenotetext{a}{This object was incorrectly identified in \citet{Kobulnicky2016}. The star's coordinates and name have been corrected here.}
\end{deluxetable}
\end{longrotatetable}
\newpage

\begin{longrotatetable}
\begin{deluxetable}{rcrrcrrrrrl}
\tablecaption{Kinematic data for bowshock stars \label{tab:kinematic}}
\tabletypesize\scriptsize 
\tablehead{
\colhead{ID} &\colhead{Name}  &\colhead{GDR2 ID}&\colhead{$V_{\odot}$ } &\colhead{parallax}  &\colhead{$\sigma$}   &\colhead{$\mu_{\alpha}$}&\colhead{$\sigma$}   &\colhead{$\mu_{\delta}$}&\colhead{$\sigma$}       \\
\colhead{}   &\colhead{    } &\colhead{    }&\colhead{(km~s$^{-1})$} &\colhead{($\mu$as)}&\colhead{($\mu$as)}&\colhead{($\mu$as~yr$^{-1}$)}    &\colhead{($\mu$as~yr$^{-1}$)}&\colhead{($\mu$as~yr$^{-1}$)}    &\colhead{($\mu$as~yr$^{-1}$)} \\
\colhead{(1)}&\colhead{(2)} &\colhead{(3)}	 &\colhead{(4)}       &\colhead{(5)}           &\colhead{(6)}       &\colhead{(7)}           &\colhead{(8)}       &\colhead{(9)} &\colhead{(10) }	     
}
\startdata 
1 & G000.1169$-$00.5703 & 4057291747277127040 & -15 & 389 & 60 & 1394 & 94 & -1045 & 73       \\
3 & G001.0563$-$00.1499 & 4057634103373474560 & -27 & 640 & 84 & -2334 & 135 & -12128 & 109   \\
7 & G003.8417$-$01.0440 & 4064254301560852352 & 0 & 736 & 47 & 2020 & 84 & -1929 & 67         \\
11 & G005.5941+00.7335 & 4067804899497964416 & 18 & 385 & 58 & 48 & 111 & 1490 & 85           \\
13 & G006.2812+23.5877 & 4337352305315545088 & -9 & 8910 & 200 & 15260 & 260 & 24790 & 220    \\
16 & G006.8933+00.0743 & 4069437463777653632 & -24 & 307 & 65 & 139 & 90 & -949 & 74          \\
26 & G011.0709$-$00.5437 & 4094704481502396288 & -1 & 298 & 42 & 358 & 78 & -888 & 64         \\
28 & G011.6548+00.4943 & 4095751942430811776 & -27 & 390 & 61 & -820 & 91 & -2544 & 78        \\
32 & G012.3407$-$00.3949 & 4095629759211238400 & 32 & 394 & 79 & -887 & 104 & -2037 & 85      \\
46 & G014.4703$-$00.6427 & 4097416946633635584 & -17 & 1486 & 244 & -4522 & 416 & -1792 & 406 \\
67 & G017.0826+00.9744 & 4146617819936287488 & 9 & 638 & 49 & -16 & 83 & -1329 & 72           \\
83 & G019.8107+00.0965 & 4153376178622946560 & 0 & 340 & 38 & -965 & 69 & -2156 & 62          \\
100 & G023.0958+00.4411 & 4156581976571298944 & 15 & 526 & 63 & 67 & 99 & -2455 & 93          \\
101 & G023.1100+00.5458 & 4156594925877520512 & 14 & 404 & 66 & 455 & 116 & -2257 & 102       \\
129 & G027.3338+00.1784 & 4256681923166288640 & -25 & 1130 & 93 & 1277 & 169 & -755 & 130     \\
163 & G031.9308+00.2676 & 4266110686393370112 & 71 & 187 & 37 & -2065 & 60 & -4389 & 54       \\
201 & G037.2933+00.6703 & 4281150081268339072 & 18 & 926 & 183 & -974 & 296 & -3618 & 262     \\
289 & G055.5792+00.6749 & 2017772886883963392 & 18 & 441 & 48 & -3204 & 55 & -4618 & 67       \\
320 & G073.6200+01.8522 & 2059236196250413696 & -4 & 371 & 35 & -2567 & 52 & -4979 & 65       \\
322 & G074.3117+01.0041 & 2060507437839909248 & -15 & 190 & 27 & -3163 & 40 & -4980 & 44      \\
329 & G077.0505$-$00.6094 & 2063868163830050176 & -39 & 441 & 54 & -3802 & 87 & -8341 & 97    \\
331 & G078.2869+00.7780 & 2067267299727875584 & -51 & 598 & 42 & -3330 & 60 & -5313 & 60      \\
333 & G078.5197+01.0652 & 2067382031192434816 & -12 & 617 & 84 & -3072 & 134 & -4506 & 159    \\
338 & G079.8223+00.0959 & 2064738049323468928 & -10 & 526 & 61 & -2357 & 103 & -4400 & 116    \\
339 & G080.2400+00.1354 & 2064838375463800448 & -43 & 557 & 40 & 4768 & 66 & 2509 & 74        \\
341 & G080.8621+00.9749 & 2067963015711183744 & -38 & 587 & 29 & -2112 & 51 & -3843 & 46      \\
342 & G080.9020+00.9828 & 2067963535403817728 & -17 & 621 & 40 & -3143 & 69 & -3058 & 68       \\
344 & G082.4100+02.3254 & 2069819545390584192 & \nodata & 606 & 30 & -2619 & 50 & 834 & 52    \\
353 & G104.3447+02.2299 & 2201205412482296448 & -76 & 224 & 27 & -6110 & 45 & -1240 & 46      \\
356 & G106.6327+00.3917 & 2008430237808841600 & -91 & 358 & 31 & -2725 & 46 & -1841 & 47      \\
357 & G106.6375+00.3783 & 2008383302396655488 & -108 & 642 & 31 & -3827 & 48 & -1691 & 47     \\
361 & G108.9891+01.5606 & 2206818556775783552 & -53 & 224 & 30 & -1875 & 42 & -981 & 39       \\
362 & G109.1157+00.6799 & 2014562519088745344 & -33 & 148 & 30 & -1620 & 53 & -1386 & 42      \\
367 & G133.1567+00.0432 & 507686819685070208 & 52 & 440 & 26 & -750 & 29 & -279 & 49           \\
368 & G134.3552+00.8182 & 465523778576137600 & -1 & 406 & 32 & -1341 & 34 & -553 & 52          \\
369 & G137.4203+01.2792 & 464697873547937664 & -49 & 331 & 47 & -142 & 53 & -299 & 76         \\
380 & G223.7092$-$01.9008 & 3046530911350220416 & 21 & 7768 & 785 & 7193 & 1377 & -7043 & 1206\\
381 & G224.1685$-$00.7784 & 3046582725837564800 & 57 & 855 & 80 & -2055 & 142 & 2645 & 169    \\
382 & G224.7096$-$01.7938 & 3046209987096803584 & 31 & 1070 & 350 & -3140 & 720 & 3320 & 550  \\
384 & G253.7104$-$00.1920 & 5543183649492723584 & \nodata & 232 & 28 & -2813 & 40 & 3273 & 44      \\
391 & G268.9550$-$01.9022 & 5325452481440825600 & \nodata & 512 & 29 & -5124 & 46 & 3393 & 45      \\
392 & G269.2089$-$00.9138 & 5326951150144920064 & \nodata & 497 & 31 & -4943 & 51 & 4739 & 53      \\
394 & G272.5794$-$01.7247 & 5313479796251809664 & \nodata & 427 & 29 & -1142 & 49 & -571 & 53      \\
395 & G273.1192$-$01.9620 & 5311906841793959936 & \nodata & 344 & 36 & -5216 & 81 & 3597 & 77      \\
397 & G282.1647$-$00.0256 & 5259053730157539328 & \nodata & 369 & 27 & -7909 & 47 & 4013 & 45       \\
404 & G286.0498$-$01.6583 & 5255179905867530752 & \nodata & 222 & 26 & -5135 & 54 & 2393 & 44       \\
405 & G286.4644$-$00.3478 & 5350640192585837568 & \nodata & 401 & 26 & -7009 & 48 & 3301 & 46      \\
406 & G287.1148$-$01.0236 & 5254478593582508288 & \nodata & 332 & 60 & -6697 & 121 & 2400 & 128    \\
407 & G287.4071$-$00.3593 & 5350395383778733568 & \nodata & 294 & 34 & -5996 & 54 & 2152 & 55      \\
409 & G287.6131$-$01.1302 & 5254268518156437888 & \nodata & 262 & 36 & -7120 & 64 & 1870 & 56       \\
410 & G287.6736$-$01.0093 & 5254269961265754368 & \nodata & 413 & 27 & -6968 & 45 & 1713 & 44      \\
411 & G288.1505$-$00.5059 & 5338310887667141888 & \nodata & 455 & 31 & -6171 & 53 & 2147 & 48      \\
413 & G288.3138$-$01.3085 & 5242187050013731200 & \nodata & 595 & 77 & -7892 & 128 & 3481 & 119    \\
414 & G288.4263$-$01.2245 & 5242184954087251456 & \nodata & 356 & 26 & -6782 & 47 & 2264 & 49      \\
463 & G308.0406+00.2473 & 5865357598905884288 & \nodata & 238 & 38 & -6063 & 41 & -1707 & 50      \\
555 & G326.7256+00.7742 & 5885668499206748160 & \nodata & 345 & 49 & -3667 & 106 & -2211 & 94     \\
589 & G332.4863+00.8256 & 5935169268617561344 & \nodata & 366 & 41 & -2564 & 82 & -2358 & 64      \\
624 & G340.4772$-$00.1528 & 5964264231754517504 &  10 & 252 & 45 & -2385 & 89 & -3248 & 61    \\
626 & G340.8579$-$00.8793 & 5964030074446095744 & -13 & 436 & 66 & -851 & 111 & -1131 & 84    \\
634 & G342.3422$-$00.4456 & 5964575672745370752 & -11 & 530 & 58 & -295 & 105 & -2863 & 89    \\
635 & G342.5873+00.1600 & 5964883879575434240 & -26 & 313 & 73 & -1166 & 148 & -818 & 99      \\
637 & G342.7172$-$00.4361 & 5964678060440055296 & -38 & 743 & 92 & -676 & 203 & -3872 & 153   \\
648 & G344.4658$-$00.5580 & 5965562308305643008 & -24 & 514 & 113 & -3205 & 197 & -3174 & 157 \\
653 & G346.1388$-$00.2184 & 5966132787371809536 & -21 & 574 & 38 & -464 & 62 & -3330 & 44     \\
655 & G346.2958+00.0744 & 5966890930670155520 & -68 & 434 & 46 & -1480 & 76 & -2738 & 53      \\
673 & G349.5431$-$00.5952 & 5972791773374952192 & -22 & 342 & 40 & -1170 & 74 & -2088 & 50    \\
692 & G353.4162+00.4482 & 5975945894262363776 & -30 & 558 & 51 & 1410 & 85 & -2094 & 59       \\
694 & G355.4972$-$00.7571 & 4053776191953892224 & -11 & 743 & 104 & 170 & 179 & -951 & 144    \\
700 & G356.6602+00.9209 & 4055297370601329024 & -25 & 271 & 46 & -392 & 71 & -1478 & 53       \\
709 & G359.9536$-$00.5088 & 4057277728502494080 & -7 & 299 & 45 & 767 & 80 & -2229 & 61       \\
\enddata
\tablecomments{(1) Identifier from \citet{Kobulnicky2016}, (2) generic identifier in Galactic coordinates, 
(3) GDR2 identifier, (4) Heliocentric radial velocity measured or adopted, as described in the Appendix, 
(5) $Gaia$ DR2 parallax in microarcseconds, (6) parallax uncertainty, 
(7) GDR2 proper motion in right ascension in microarcseconds per year, (8) uncertainty, 
(9) GDR2 proper motion in declination in microarcseconds per year, (10) uncertainty.  }
\end{deluxetable}
\end{longrotatetable}
\newpage

\begin{longrotatetable}
\begin{deluxetable}{rcrcrrrrrrrrrr}
\tablecaption{Derived parameters for stars and their bowshock nebulae \label{tab:derived}}
\movetabledown=0.5 in
\tabletypesize\scriptsize 
\tablehead{
\colhead{ID} &\colhead{Name}&\colhead{Alt. name}&\colhead{Sp.T.}&\colhead{Lum.}              &\colhead{$U$}&\colhead{$j_\nu$}           &\colhead{$f_{j}$}     &\colhead{$n_{\rm a}$}                &\colhead{$V_{\rm tot}$}    &\colhead{$\sigma_{V_{\rm tot}}$}    &\colhead{$\dot M$}  &\colhead{$\sigma$}  &\colhead{$\Delta\log(\dot M)$ } \\
\colhead{}   &\colhead{    }&\colhead{         }&\colhead{     }&\colhead{(10$^4$ L$_\odot$)}&\colhead{(10$^{2}$)}   &\colhead{(10$^{-13}$~Jy~sr$^{-1}$~cm$^2$)}&\colhead{} &\colhead{(cm$^{-3}$)} &\colhead{(km~s$^{-1}$)}  &\colhead{(km~s$^{-1}$)}  & \colhead{(10$^{-10}$ M$_\odot$~yr$^{-1}$)} &\colhead{(10$^{-10}$ M$_\odot$~yr$^{-1}$)}  &\colhead{} \\
\colhead{(1)}&\colhead{(2)} &\colhead{(3)}	&\colhead{(4)}  &\colhead{(5)}               &\colhead{(6)}&\colhead{(7)}                  &\colhead{(8)}        &\colhead{(9)}	 &\colhead{(10)}  &\colhead{(11)} &\colhead{(12)} &\colhead{(13)}	&\colhead{(14)}	 }
\startdata 
1 & \nodata & G000.1169$-$00.5703 & O8V & 12.0 & 16 & 73 & 2.03 & 41 & 16.3 & 3.8 & 2639 & 1109 & 0.06 \\
3 & \nodata & G001.0563$-$00.1499 & B5III: & 0.28 & 18 & 76 & 2.02 & 530 & 75.6 & 14.3 & 62957 & 33928 & 4.04 \\
7 & \nodata & G003.8417$-$01.0440 & B2V & 0.46 & 20 & 81 & 1.99 & 453 & 12.1 & 1.3 & 1411 & 587 & 2.03 \\
11 & \nodata & G005.5941+00.7335 & B0V & 3.4 & 56 & 135 & 1.8 & 7 & 31.5 & 3.8 & 201 & 88 & 0.19 \\
13 & Zeta Oph & G006.2812+23.5877 & O9.2IV & 4.2 & 21 & 83 & 1.98 & 9 & 11.9 & 0.1 & 117 & 45 & -0.31 \\
16 & \nodata & G006.8933+00.0743 & B0III & 14.0 & 10 & 57 & 2.15 & 37 & 5.0 & 3.2 & 673 & 319 & -0.56 \\
26 & \nodata & G011.0709$-$00.5437 & O9V & 5.1 & 46 & 123 & 1.84 & 17 & 8.2 & 2.5 & 67 & 27 & -0.79 \\
28 & \nodata & G011.6548+00.4943 & B0V & 3.4 & 39 & 114 & 1.86 & 9 & 15.8 & 4.4 & 96 & 38 & -0.14 \\
32 & \nodata & G012.3407$-$00.3949 & B2V+B2V & 0.46 & 8 & 49 & 2.21 & 73 & 11.5 & 4.7 & 227 & 98 & 1.23 \\
46 & \nodata & G014.4703$-$00.6427 & B2V: & 0.46 & 25 & 91 & 1.95 & 33 & 17.5 & 3.9 & 169 & 77 & 1.11 \\
67 & NGC6611ESL45 & G017.0826+00.9744 & O9V & 5.1 & 229 & 255 & 1.63 & 163 & 5.5 & 0.9 & 56 & 25 & -0.87 \\
83 & \nodata & G019.8107+00.0965 & B8II & 0.29 & 7 & 46 & 2.25 & 63 & 9.3 & 1.4 & 208 & 129 & 2.28 \\
100 & \nodata & G023.0958+00.4411 & B1V & 1.6 & 63 & 143 & 1.79 & 23 & 5.0 & 1.9 & 12 & 5 & -1.19 \\
101 & \nodata & G023.1100+00.5458 & B0V & 3.4 & 22 & 85 & 1.98 & 51 & 9.0 & 3.7 & 352 & 160 & 0.43 \\
129 & TYC5121-625-1 & G027.3338+00.1784 & M0III & 0.029 & 1 & 19 & 2.68 & 2111 & 11.3 & 1.2 & \nodata & \nodata & \nodata \\
163 & \nodata & G031.9308+00.2676 & B3II & 0.88 & 1 & 14 & 2.85 & 51 & 18.2 & 9.8 & 15376 & 8977 & 2.68 \\
201 & \nodata & G037.2933+00.6703 & B2V & 0.46 & 38 & 112 & 1.87 & 40 & 5.0 & 2.7 & 9 & 4 & -0.13 \\
289 & \nodata & G055.5792+00.6749 & M0III & 0.029 & 0 & 0 & 4.48 & 981 & 16.6 & 1.2 & \nodata & \nodata & \nodata \\
320 & HD191611 & G073.6200+01.8522 & O9III & 14.0 & 10 & 56 & 2.15 & 31 & 5.0 & 1.3 & 548 & 195 & -0.66 \\
322 & \nodata & G074.3117+01.0041 & O8IV: & 19.0 & 14 & 68 & 2.07 & 14 & 11.4 & 2.4 & 754 & 306 & -0.80 \\
329 & KGK2010-10 & G077.0505$-$00.6094 & O8V & 7.9 & 101 & 177 & 1.72 & 11 & 41.6 & 5.2 & 474 & 185 & -0.32 \\
331 & LSII+39~53 & G078.2869+00.7780 & O7V & 12.0 & 15 & 70 & 2.06 & 14 & 12.9 & 0.7 & 471 & 206 & -0.69 \\
333 & \nodata & G078.5197+01.0652 & B0V+B0V & 4.4 & 12 & 62 & 2.11 & 74 & 8.8 & 1.6 & 270 & 109 & -0.03 \\
338 & CPR2002A10 & G079.8223+00.0959 & O8V: & 7.9 & 26 & 93 & 1.94 & 114 & 5.0 & 1.8 & 282 & 115 & -0.54 \\
339 & CPR2002A37 & G080.2400+00.1354 & O5V & 32.0 & 12 & 63 & 2.1 & 35 & 78.1 & 6.6 & 121619 & 49228 & 0.90 \\
341 & KGK2010-1 & G080.8621+00.9749 & O9V & 5.1 & 27 & 95 & 1.93 & 8 & 5.0 & 0.8 & 17 & 6 & -1.39 \\
342 & KGK2010-2 & G080.9020+00.9828 & B4V & 0.093 & 2 & 23 & 2.59 & 795 & 12.2 & 1.1 & 4115 & 2115 & 3.86 \\
344 & BD+43~3654 & G082.4100+02.3254 & O4If & 87.0 & 5 & 39 & 2.33 & 39 & 37.7 & 2.3 & 197923 & 72806 & 0.42 \\
353 & \nodata & G104.3447+02.2299 & O5V & 32.0 & 12 & 63 & 2.1 & 7 & 65.2 & 7.9 & 17007 & 6931 & 0.05 \\
356 & HD240015 & G106.6327+00.3917 & B0III & 14.0 & 49 & 127 & 1.83 & 5 & 5.0 & 1.0 & 22 & 9 & -2.05 \\
357 & HD240016 & G106.6375+00.3783 & B2III & 3.8 & 20 & 81 & 1.99 & 8 & 13.5 & 0.5 & 245 & 93 & -0.41 \\
361 & TYC4278-522-1 & G108.9891+01.5606 & B1Ia & 8.7 & 0 & 7 & 3.25 & 12 & 19.2 & 2.1 & 76045 & 45397 & 1.43 \\
362 & \nodata & G109.1157+00.6799 & B2I & 13.0 & 12 & 63 & 2.1 & 4 & 22.4 & 3.6 & 1608 & 721 & -0.62 \\
367 & LSI+60~226 & G133.1567+00.0432 & O7V & 12.0 & 11 & 61 & 2.12 & 8 & 5.0 & 1.0 & 50 & 20 & -1.66 \\
368 & KM~Cass & G134.3552+00.8182 & O8.5V & 7.0 & 39 & 113 & 1.87 & 31 & 5.4 & 1.3 & 53 & 22 & -1.12 \\
369 & BD+60~586 & G137.4203+01.2792 & O8III & 19.0 & 2 & 25 & 2.54 & 20 & 8.4 & 3.1 & 3278 & 1549 & -0.17 \\
380 & HD53367 & G223.7092$-$01.9008 & B2Ve: & 0.46 & 322 & 297 & 1.6 & 131 & 10.7 & 1.5 & 20 & 7 & 0.46 \\
381 & HD54662 & G224.1685$-$00.7784 & O7Vz & 12.0 & 1 & 14 & 2.84 & 10 & 6.8 & 1.5 & 1287 & 530 & -0.25 \\
382 & FN~Cma & G224.7096$-$01.7938 & B0III & 12.0 & 7 & 48 & 2.22 & 25 & 9.1 & 5.2 & 1764 & 1968 & 0.07 \\
384 & CD-35~4415 & G253.7104$-$00.1920 & O7IIInn & 26.0 & 10 & 58 & 2.14 & 32 & 5.1 & 3.2 & 530 & 203 & -1.25 \\
391 & HD77207 & G268.9550$-$01.9022 & B7Iab + & 3.2 & 5 & 38 & 2.33 & 15 & 11.2 & 1.1 & 1538 & 940 & 0.85 \\
392 & CPD-47~3051 & G269.2089$-$00.9138 & B0V & 3.4 & 6 & 45 & 2.26 & 12 & 5.0 & 1.2 & 77 & 33 & -0.23 \\
394 & HD298310 & G272.5794$-$01.7247 & B0V & 14.0 & 37 & 111 & 1.87 & 5 & 67.0 & 4.1 & 3914 & 1446 & 0.20 \\
395 & HD298353 & G273.1192$-$01.9620 & O7V & 12.0 & 9 & 54 & 2.17 & 18 & 10.1 & 2.7 & 604 & 271 & -0.58 \\
397 & CD-55~3196 & G282.1647$-$00.0256 & O9.5III & 13.0 & 24 & 89 & 1.96 & 28 & 23.2 & 2.9 & 4378 & 1787 & 0.28 \\
404 & CD-59~3123 & G286.0498$-$01.6583 & O9.5Ib & 30.0 & 3 & 31 & 2.43 & 7 & 24.4 & 6.4 & 16560 & 7927 & 0.37 \\
405 & HD303197 & G286.4644$-$00.3478 & B3III & 0.88 & 2 & 22 & 2.61 & 62 & 7.7 & 1.5 & 1706 & 821 & 1.72 \\
406 & HD92607 & G287.1148$-$01.0236 & O9IVn & 8.2 & 24 & 88 & 1.96 & 32 & 7.9 & 4.9 & 338 & 126 & -0.46 \\
407 & HD93249 & G287.4071$-$00.3593 & O9III+O & 14.0 & 141 & 207 & 1.68 & 26 & 9.1 & 2.1 & 109 & 43 & -1.37 \\
409 & HD93027 & G287.6131$-$01.1302 & O9.5IV & 4.0 & 36 & 108 & 1.88 & 23 & 22.1 & 5.2 & 693 & 284 & 0.44 \\
410 & HD305536 & G287.6736$-$01.0093 & O9.5V+ B & 4.1 & 338 & 302 & 1.6 & 66 & 11.6 & 1.6 & 30 & 11 & -0.95 \\
411 & HD305599 & G288.1505$-$00.5059 & O9.5V & 4.1 & 314 & 294 & 1.61 & 79 & 6.8 & 1.0 & 26 & 11 & -1.01 \\
413 & HD93683 & G288.3138$-$01.3085 & O9V+B0V & 5.1 & 49 & 126 & 1.83 & 19 & 9.2 & 3.5 & 41 & 16 & -1.00 \\
414 & HD93858 & G288.4263$-$01.2245 & B3II & 0.88 & 22 & 85 & 1.98 & 16 & 6.0 & 1.6 & 25 & 12 & -0.11 \\
463 & TYC8995-1548-1 & G308.0406+00.2473 & B0III: & 12.0 & 20 & 80 & 2.0 & 9 & 5.0 & 2.7 & 67 & 27 & -1.35 \\
555 & ALS18049 & G326.7256+00.7742 & O9V: & 5.1 & 243 & 263 & 1.63 & 82 & 13.0 & 2.2 & 150 & 57 & -0.44 \\
589 & CD-49 10393 & G332.4863+00.8256 & B0V & 3.4 & 33 & 105 & 1.89 & 10 & 7.4 & 1.5 & 27 & 11 & -0.67 \\
624 & \nodata & G340.4772$-$00.1528 & B0III & 14.0 & 19 & 78 & 2.01 & 9 & 27.4 & 5.0 & 2496 & 1039 & 0.01 \\
626 & \nodata & G340.8579$-$00.8793 & O9V & 5.1 & 47 & 124 & 1.83 & 187 & 12.8 & 1.5 & 1717 & 645 & 0.62 \\
634 & HD152756 & G342.3422$-$00.4456 & B0III & 12.0 & 20 & 82 & 1.99 & 10 & 5.9 & 2.4 & 101 & 42 & -1.17 \\
635 & \nodata & G342.5873+00.1600 & O6V & 19.0 & 11 & 59 & 2.13 & 20 & 16.7 & 3.7 & 2244 & 1458 & -0.38 \\
637 & \nodata & G342.7172$-$00.4361 & B2V & 0.46 & 45 & 121 & 1.84 & 13 & 5.0 & 3.5 & 3 & 1 & -0.63 \\
648 & \nodata & G344.4658$-$00.5580 & B0III & 14.0 & 102 & 177 & 1.72 & 8 & 23.7 & 10.7 & 339 & 130 & -0.86 \\
653 & \nodata & G346.1388$-$00.2184 & K0III: & 0.005 & 2 & 26 & 2.51 & 87 & 7.4 & 1.6 & \nodata & \nodata & \nodata \\
655 & \nodata & G346.2958+00.0744 & B1V: & 1.6 & 163 & 221 & 1.66 & 19 & 12.3 & 3.0 & 23 & 9 & -0.91 \\
673 & \nodata & G349.5431$-$00.5952 & B1V: & 1.6 & 35 & 107 & 1.88 & 10 & 12.9 & 3.1 & 59 & 23 & -0.5 \\
692 & \nodata & G353.4162+00.4482 & O7V & 12.0 & 20 & 81 & 2.0 & 116 & 13.6 & 2.2 & 3416 & 1638 & 0.17 \\
694 & \nodata & G355.4972$-$00.7571 & B2V: & 0.49 & 46 & 124 & 1.83 & 17 & 9.5 & 1.8 & 14 & 7 & -0.03 \\
700 & \nodata & G356.6602+00.9209 & B0V & 3.4 & 6 & 45 & 2.26 & 8 & 9.5 & 4.8 & 180 & 84 & 0.14 \\
709 & \nodata & G359.9536$-$00.5088 & B0V+B1V & 3.4 & 16 & 72 & 2.04 & 12 & 21.5 & 6.0 & 317 & 127 & 0.38 \\
\enddata
\tablecomments{(1) Identifier from \citet{Kobulnicky2016}, (2) Common name, (3) generic identifier in Galactic coordinates, (4) spectral classification, (5)
stellar luminosity in units of 10$^4$ times the solar luminosity, computed from effective temperature and radius in Table~\ref{tab:basic}, (6) dimensionless ratio of the radiant energy density (in erg~cm$^{-3}$) from to the star to the mean interstellar radiant energy density estimated by \citet[][MMP83]{Mathis1983} (7) dust emission coefficient in 10$^{-13}$ Jy sr$^{-1}$ cm$^2$ {\it per nucleon}, following DL07, after scaling by the factor $f_j$,(8) scale factor, $f_j$, for dust emission coefficients on account of the harder radiation field near OB stars, as described in the text, (9) computed ambient interstellar number density, following Equation 5 of \citet{Kobulnicky2018}, (10) peculiar velocity in the star's local standard of rest; an arbitrary minimum of 5 \kms\ has been imposed.  Velocities include only the two plane-of-sky components and not the radial component, as discussed in the text.
(11) peculiar velocity uncertainty, (12) computed mass-loss rate in 10$^{-10}$ solar masses per year, (13) uncertainty on mass-loss rate. Uncertainties do not include the uncertainties on the star's peculiar velocity and, hence, are lower limits, (14) the base-10 logarithm of our derived mass-loss rate minus the logarithm of the theoretical \citet{Vink2001} mass-loss rate, after correcting the theoretical values downward by 0.12 dex on account revision to the solar abundance scale since that work. } 
\tablenotetext{a}{For G012.3407$-$00.3949, G078.5197+01.0652, G287.4071$-$00.3593, G288.3138$-$01.3085, G359.9536$-$00.5088 we have divided the mass loss rate by a factor of two in order to account for the double nature of these sources. There are certain to be other unidentified binaries among our sample. } 
\end{deluxetable}
\end{longrotatetable}

\newpage

\begin{deluxetable}{ll}
\tablecaption{Objects in special categories \label{tab:classifications}}
\tabletypesize\scriptsize 
\tablehead{
\colhead{Category} &\colhead{ID}     
}
\startdata 
$V_{\rm tot}<$ 5~km~s$^{-1}$    &  16, 100, 201, 320, 338, 341, 356, 367, 392, 463, 637 \\
Faces 8 $\mu$m bright-rimmed cloud & 3, 7, 380, 384, 101, 555, 709 \\
Bow wave candidate, $L_*/L_{\rm IR}<75$  & 3, 7, 32, 46, 67, 101, 129, 163, 289, 243, 405, 555, 626, 653, 709
\enddata
\tablecomments{(1) Object classification denoting special object categories according to the criteria discussed in Section~4, (2) identification numbers of objects in each category.   } 
\end{deluxetable}

\newpage

\newpage

\newpage

\newpage

\newpage

\newpage

\newpage

\newpage

\newpage

\newpage

\newpage

\newpage

\newpage

\newpage

\end{document}